\documentclass[modern,astrosymb]{aastex631}

\received{\today}
\revised{\today}
\accepted{\today}

\newcommand{\teff}{$T_\mathrm{eff}$}
\newcommand{\lbol}{$L_\mathrm{bol}$}

\newcommand{\rjup}{$R_\mathrm{Jup}$}

\submitjournal{\apj}

\shorttitle{W0146}
\shortauthors{Bardalez Gagliuffi et al.}
\graphicspath{{./}{figures/}}

\usepackage{graphics}
\usepackage{multirow}
\usepackage{hyperref}

\begin{document}

\title{The Diversity of Cold Worlds: a blended-light binary straddling the T/Y transition in brown dwarfs}

\correspondingauthor{Daniella C. Bardalez Gagliuffi}
\email{dbardalezgagliuffi@amherst.edu}

\author[0000-0001-8170-7072]{Daniella C. Bardalez Gagliuffi}
\affiliation{Department of Physics \& Astronomy, Amherst College, Amherst, MA 01002, USA}
\affiliation{Department of Astrophysics, American Museum of Natural History, New York, NY, USA}

\author[0000-0001-6251-0573]{Jacqueline K. Faherty}
\affiliation{Department of Astrophysics, American Museum of Natural History, New York, NY, USA}

\author[0000-0002-2011-4924]{Genaro Su\'{a}rez}
\affiliation{Department of Astrophysics, American Museum of Natural History, New York, NY, USA}

\author[0000-0003-0548-0093]{Sherelyn Alejandro Merchan}
\affiliation{Department of Astrophysics, American Museum of Natural History, New York, NY, USA}
\affiliation{Department of Physics, The Graduate Center City University of New York, New York, NY, USA.}

\author[0000-0002-9420-4455]{Brianna Lacy}
\affiliation{Department of Astronomy, University of Texas at Austin, Austin, TX, USA}
\affiliation{Department of Astronomy and Astrophysics, University of California, Santa Cruz, CA, USA}

\author[0000-0003-4600-5627]{Ben Burningham}
\affiliation{Department of Physics, Astronomy and Mathematics, University of Hertfordshire, Hatfield, UK}

\author[0009-0007-3572-0664]{Klara Matuszewska}
\affiliation{Department of Physics \& Astronomy, Amherst College, Amherst, MA, USA}

\author[0000-0003-2102-3159]{Rocio Kiman}
\affiliation{Department of Astronomy, California Institute of Technology, Pasadena, CA, USA}

\author[0000-0003-0489-1528]{Johanna M. Vos}
\affiliation{School of Physics, Trinity College Dublin, The University of Dublin, Dublin, Ireland}

\author[0000-0003-4083-9962]{Austin Rothermich}
\affiliation{Department of Astrophysics, American Museum of Natural History, New York, NY, USA}
\affiliation{Department of Physics, The Graduate Center City University of New York, New York, NY, USA}

\author[0000-0002-2592-9612]{Jonathan Gagn\'{e}}
\affiliation{Plan\'{e}tarium Rio Tinto Alcan, Montreal, Quebec, Canada}
\affiliation{D\'{e}partement de Physique, Universit\'{e} de Montr\'{e}al, Montreal, Quebec, Canada}

\author[0000-0002-4404-0456]{Caroline Morley}
\affiliation{Department of Astronomy, University of Texas at Austin, Austin, TX, USA.}

\author[0000-0003-4225-6314]{Melanie J. Rowland}
\affiliation{Department of Astronomy, University of Texas at Austin, Austin, TX, USA.}

\author[0000-0001-7896-5791]{Dan Caselden}
\affiliation{Department of Astrophysics, American Museum of Natural History, New York, NY, USA}

\author[0000-0002-1125-7384]{Aaron Meisner}
\affiliation{NSF’s National Optical-Infrared Astronomy Research Laboratory, Tucson, AZ, USA}

\author[0000-0002-6294-5937]{Adam C. Schneider}
\affiliation{United States Naval Observatory, Flagstaff, AZ, USA}

\author[0000-0002-2387-5489]{Marc J. Kuchner}
\affiliation{Exoplanets and Stellar Astrophysics Laboratory, NASA
Goddard Space Flight Center, Greenbelt, MD, USA}

\author[0000-0002-5627-5471]{Charles A. Beichman}
\affiliation{IPAC, Caltech, Pasadena, CA, USA}

\author{Peter R. Eisenhardt}
\affiliation{NASA Jet Propulsion Laboratory, California Institute of Technology, Pasadena, CA, USA}

\author[0000-0001-5072-4574]{Christopher R. Gelino}
\affiliation{IPAC, Caltech, Pasadena, CA, USA}

\author[0000-0002-4088-7262]{Ehsan Gharib-Nezhad}
\affiliation{NASA Ames Research Center, Mountain View, CA, USA}

\author[0000-0003-4636-6676]{Eileen C. Gonzales}
\affil{Department of Physics and Astronomy, San Francisco State University San Francisco, CA, USA}

\author[0000-0001-7519-1700]{Federico Marocco}
\affiliation{IPAC, Caltech, Pasadena, CA, USA}

\author[0000-0001-8818-1544]{Niall Whiteford}
\affiliation{Department of Astrophysics, American Museum of Natural History, New York, NY, USA}

\author[0000-0003-4269-260X]{J. Davy Kirkpatrick}
\affiliation{IPAC, Caltech, Pasadena, CA, USA}

\begin{abstract}
We present the first brown dwarf spectral binary characterized with JWST: WISE J014656.66+423410.0, the coldest blended-light brown dwarf binary straddling the T/Y transition. We obtained a moderate resolution (R$\sim$2700) G395H spectrum of this unresolved binary with JWST/NIRSpec and we fit it to late-T and Y dwarf spectra from JWST/NIRSpec, and model spectra of comparable temperatures, both as individual spectra and pairs mimicking an unresolved binary system. We find that this tightly-separated binary is likely composed of two unequal-brightness sources with a magnitude difference of $0.50\pm0.08$ mag in IRAC [4.5] and a secondary $1.01\pm0.13$ mag redder than the primary in [3.6]-[4.5]. Despite the large color difference between the best fit primary and secondary, their temperature difference is only $92\pm23$\,K, a feature reminiscing of the L/T transition. Carbon disequilibrium chemistry strongly shapes the mid-infrared spectra of these sources, as a complex function of metallicity and surface gravity. While a larger library of JWST/NIRSpec spectra is needed to conclusively examine the peculiarities of blended-light sources, this spectral binary is a crucial pathfinder to both understand the spectral features of planetary-mass atmospheres and detect binarity in unresolved, moderate-resolution spectra of the coldest brown dwarfs.
\end{abstract}


\section{Introduction}\label{sec:intro}

With temperatures below 500\,K, Y dwarfs are the coldest brown dwarfs, which mark the low-mass end of the substellar classification sequence~\citep{2011ApJ...743...50C}. Due to their inability to fuse hydrogen, brown dwarfs cool over time across spectral types late-M, L, T, and Y. Brown dwarfs are classified as Y based on their near-infrared (near-IR) spectral morphology, and can either be very young, very low-mass objects or older, more massive ones, with about 63\% of them having ``planetary'' masses below the deuterium burning limit based on evolutionary models~\citep{2019ApJ...882L..29M}. Their planet-like characteristics such as mass, temperature, surface gravity, and atmosphere make these objects the closest analogs to cold, long-period giant exoplanets and Jupiter~\citep{2016ApJ...826L..17S,2021ApJ...922L..43B,2024Natur.633..789M}.

However, little is known about their binary properties as a population. The multiplicity fraction of stellar populations decreases with primary stellar mass~\citep{2010ApJS..190....1R} and this trend continues across the hydrogen burning limit into the brown dwarf regime {
~\citep{2018MNRAS.479.2702F,2007prpl.conf..427B}. The binary properties of the coldest population of brown dwarfs remain largely unexplored, both due to their extreme faintness in the 
near-IR, and the intrinsic rarity of binaries with very low primary masses~\citep{2016ApJ...819...17O, 2018MNRAS.479.2702F}. The exceptional spatial resolution and mid-infrared (mid-IR) sensitivity of JWST/
NIRCam~\citep{2023PASP..135f8001G,2023PASP..135b8001R} were key to 
resolve WISE~J033605.05-014350.4 (hereafter, W0336AB), the first Y+Y dwarf binary, through Kernel Phase Interferometry using the F150W and F480M filters~\citep{2023ApJ...947L..30C}. For most sources, in the absence of such a dedicated observing strategy, identifying peculiarities in the unresolved spectra of binary systems can be an important avenue for the detection and characterization of brown dwarf binary systems with extremely low luminosity. 

Very low mass binary systems also tend to be closely separated, a trend that we see across stellar populations as the primary mass decreases, with characteristic binary separations of $\sim40$\,au, $\sim30$\,au, and $\sim4-7\,$au, for solar-type, M dwarf, and ultracool binaries, respectively~\citep{2022arXiv220310066O,2019AJ....157..216W,2003ApJ...587..407C,2019BAAS...51c.285B}. Moreover, while the mass ratio distribution tends to peak at unity for ultracool binaries~\citep{2007ApJ...668..492A,2004ApJS..155..191B}, the recently discovered Y+Y binary 0336AB) has an estimated mass ratio of $q=0.6\pm0.05$~\citep{2023ApJ...947L..30C}. However, this mass ratio was estimated from flux ratios in F150W and F480M, wavelength ranges which might be subject to gas or condensate absorption, impacting the mass ratio estimate.

Therefore, the very nature of low-mass multiples makes it challenging to resolve and characterize tight, faint binary systems with a large flux difference in the infrared. However, ultracool dwarf systems whose components have different temperatures and spectral features can be identified and characterized empirically through spectral peculiarities and spectral binary template fits. This \textit{spectral binary} technique has been developed and applied on unresolved, low-resolution (R$\sim$100), near-IR spectra of systems with a late-M or L-type brown dwarf primary and a T-dwarf secondary, leading to over two dozen binary discoveries~\citep{2014ApJ...794..143B,2010ApJ...710.1142B} at lower mass ratios than accessible by imaging. The key feature in the blended-light spectra of these systems is a methane absorption feature at 1.62\,$\mu$m originating from the colder T dwarf companion, yet prominent in the combined spectrum whose overall shape resembles that of the warmer and significantly brighter primary. In L/T transition systems where both late-L and early-T components are almost equally bright, this feature is even more prominent due to the near-equal brightness of late-L and early-T brown dwarfs despite their cooling trend. This is caused by the J-band brightening following the sinking or break up of clouds above the photosphere of L-dwarfs~\citep{2010ApJ...710.1142B, 2004AJ....127.3553K,1996AandA...305L...1T}.

At later types and colder temperatures, the differences in atmospheric chemistry are much more subtle. In the near-IR, late-T dwarfs show deep absorption bands of water and methane at 1.2\,$\mu$m and 1.6\,$\mu$m, which become broader at lower temperatures \citep{2006ApJ...637.1067B}, having the effect of narrowing the emission peaks in J and H bands. Therefore, the combination of early-T and late-T dwarfs does not yield peculiar absorption features in the near-IR but rather a slight difference in the width of the emission peaks. It was surmised that the onset of ammonia would mark the transition from the T to the Y spectral type but the near-IR absorption of the molecule, expected at \teff$\lesssim700$\,K between 1.53 to 1.58\,$\mu$m \citep{2011ApJ...743...50C}, was not securely detected~\citep{2015ApJ...804...92S,2024arXiv240708518B}. Therefore, the original spectral typing scheme for Y dwarfs relied on the shape of the near-IR CH$_4$ absorption feature~\citep{2011ApJ...743...50C}. However, ammonia can appear in the near-IR at 3\,$\mu$m among the coldest Y dwarfs~\citep{2024Natur.628..511F}, and in the mid-IR in objects as warm as 1200\,K at $10-11.2\,\mu$m, several hundred degrees warmer than the nominal T/Y transition temperature ($\sim500\,$K) and strengthening towards colder temperatures \citep{2022MNRAS.513.5701S}. For brown dwarfs colder than 500\,K (although also present in the near-IR throughout the T dwarf class), the $2.8-5.2\,\mu$m wavelength range covered by JWST/
NIRSpec~\citep{2022A&A...661A..80J} is severely affected by carbon disequilibrium chemistry~\citep{2020AJ....160...63M,2024arXiv240708518B}.  From ground-based observations, these atmospheres reveal enhancements of 10$^{11}$ times as much CO by mole fraction as predicted by equilibrium chemistry for a 500\,K object, where the dominant carbon molecule should be CH$_4$~\citep{2020AJ....160...63M,2014ApJ...797...41Z}. Therefore, a sudden change in the strength of carbon disequilibrium chemistry or a significant departure from equilibrium is likely to be an important marker of the T/Y transition in the mid-IR. 


In this paper, we explore the binary signatures in the blended-light, moderate-resolution JWST/
NIRSpec spectrum of WISE J014656.66+423410.0 (hereafter, W0146), a confirmed binary system whose components straddle the T/Y transition. Prior to JWST, the two best-fit components modeling this binary system were unusually faint. With a new Spitzer parallax~\citep{2021ApJS..253....7K}, our blended-light binary system indeed looks overluminous in the color-magnitude diagram (CMD; see Figure~\ref{fig:cmd}). Moreover, with our JWST/NIRSpec unresolved spectra at 3-5\,$\mu$m, we learned that the two components are not identical in \teff, and that their atmospheres have different disequilibrium chemistry strengths. Section~\ref{sec:system} provides background information on this system; Section~\ref{sec:obs} describes our JWST observations. In Section~\ref{sec:splib}, we describe building the spectral libraries for both single and binary templates and spectral models and the mechanics of spectral fitting. In Section~\ref{sec:results}, we present our results for the best single and binary fits, and in Section~\ref{sec:discussion} we discuss the signatures of blended-light binarity at these cold temperatures and the implications for identifying binary systems with JWST spectra. We present our conclusions in Section~\ref{sec:conclusions}.

\section{W0146 blended-light binary system}\label{sec:system}

\begin{figure}
\includegraphics[width=\linewidth, trim=110 50 150 20, clip]{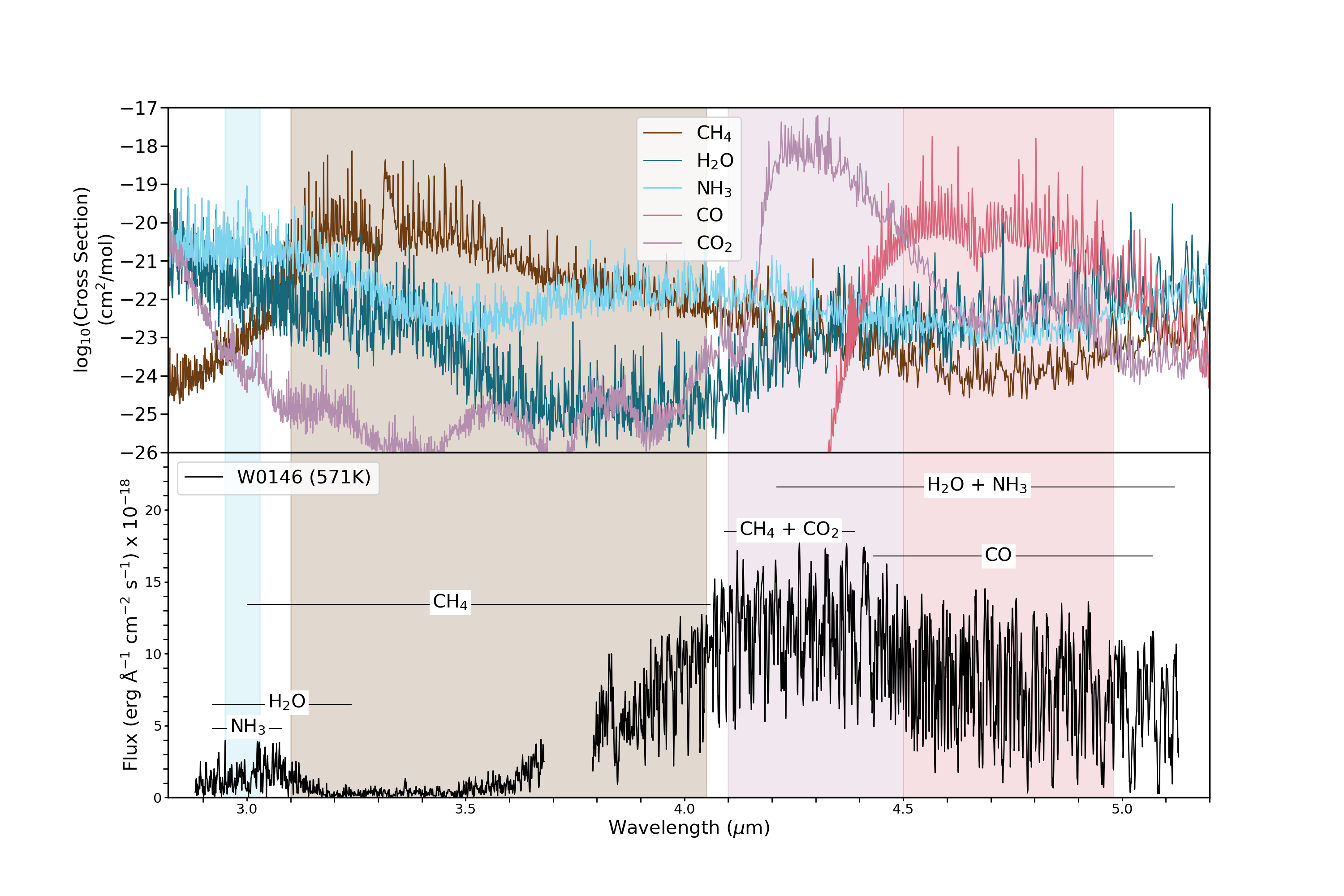}
\caption{\emph{(Top)} Cross sections of molecular species potentially occurring in Y dwarfs, calculated at a temperature of 575\,K to approximate the combined-light temperature of W0146AB. \emph{(Bottom)} The JWST/
NIRSpec spectrum of the blended-light binary W0146AB shown in black, with noise in light gray. Prominent molecular absorption features are indicated at the top. Shaded regions align between the strongest absorber at a given wavelength range and the specific absorption feature in the spectrum. \label{fig:W0146_spectrum}}
\end{figure}

W0146 was originally discovered by~\citet{2012ApJ...753..156K} and classified as a Y0 dwarf within 10\,pc. This source was partially resolved as a tightly separated binary system of T9 and Y0 components using laser guide star adaptive optics (LGS-AO) imaging with Keck/
NIRC2~\citep{2015ApJ...803..102D}. 
At the time of its discovery, the system's separation was estimated to be $0\farcs0875\pm0\farcs0021$ or $0.93^{+0.12}_{-0.16}$\,au at a parallactic distance of $10.6^{+1.3}_{-1.8}$\,pc and with over 1\,mag flux difference in $J$-band, $H$-band, and the narrow CH$_{4S}$ filter~\citep{2014ApJ...783...68B}. However, the distance to the blended-light system has been revised with more recent Spitzer astrometry to $d = 19.34\pm0.75$\,pc ($\pi = 51.7\pm2.0$\,mas;~\citealt{2021ApJS..253....7K}) thus updating the projected separation of the system to $1.69\pm0.106$\,au.

The most peculiar characteristics of this system at the time of discovery were the inferred faintness and low temperatures of its components compared to other objects of the same spectral types. The component spectral types were inferred from a blended-light, low-resolution ($R\approx750$), near-IR spectrum obtained with Gemini/
GNIRS \citep{2015ApJ...803..102D}. Combined-light photometry obtained with Gemini/
NIRI in YJHK and CH$_{4S}$ bands was used to flux calibrate the spectrum. With that constraint,~\citet{2015ApJ...803..102D} concluded that the unresolved spectrum was best reproduced by combining the T9 standard UGPS J072227.51$-$054031.2 (hereafter: U0722;~\citealt{2010MNRAS.408L..56L,2012ApJ...748...74L}) and the Y0.5pec dwarf WISEPC J140518.40+553421.5 (hereafter: W1405;~\citealt{2011ApJ...743...50C}) with the caveat that both components of this binary system needed to be $\gtrsim100$\,K colder (\teff = 345\,K and 330\,K, respectively) than the bulk of the late-T/Y dwarf population for the binary fit to work. The authors considered the possibility of an overestimated parallax measurement and calculated that the system would need to have a parallax of $\approx40$\,mas ($d = 26^{+8}_{-6}$\,pc) for the components to have typical temperatures and brightness for T9 and Y0 dwarfs based on their bolometric luminosity~\citep{2013Sci...341.1492D}. With the revision of the parallax to $\pi = 51.7\pm2.0$\,mas~\citep{2021ApJS..253....7K}, the condition of unusually faint components is no longer needed to explain the unresolved magnitude of this source.

\section{Observations and Flux Calibration}

\subsection{JWST/NIRSpec Medium Resolution Spectroscopy}~\label{sec:obs}

We have obtained a medium resolution ($R\sim2700$) spectrum of the blended-light system W0146 with JWST/
NIRSpec fixed slit spectroscopy using the G395H grating as part of Cycle 1 program GO 2124 (PI: J. Faherty). The source was observed by JWST on 2023-01-03 at 05:27:23 (UTC) following a 3-POINT-NOD dither pattern for a total exposure time of 1729.38\,s (28.8\,min). 
Data was acquired through the S200A1 fixed slit aperture matched with the SUBS200A1 subarray and using the NRSRAPID readout pattern, the clear F290LP filter, and the G395H prism grating. The spectra were reduced with the JWST calibration pipeline starting from uncalibrated data and using default parameters but optimizing the aperture extraction, as described in \citet{2024Natur.628..511F}. Only one trace appears in the 2D spectral image. Our spectrum spans the wavelength range $2.87-5.14\,\mu$m with a gap between $3.69-3.79\,\mu$m characteristic of this fixed slit mode which spans both 
NIRSpec detectors (NRS1 and NRS2). The SNR peaks at 64 in NRS2, with a median of 2.17 in NRS1 and 29.00 in NRS2. We used the JWST reduction pipeline version 1.10.0 based on the Calibration Reference Data System (CRDS) context file jwst\_1146.pmap.

\subsection{JWST/MIRI photometry}

In addition, we obtained MIRI photometry of the target in the F1000W, F1280W and F1800W filters. Observations were taken on 2022-09-21 between 01:38:51 and 01:48:56 UTC with 7, 8, and 10 groups per integration, respectively, and one integration per exposure with two dither positions. The total exposure time was 38.85\,s, 44.4\,s, and 55.5\,s for the F1000W, F1280W, and F1800W filters, respectively. We used the MAST-produced pipeline reductions of MIRI photometry, choosing the APER\_TOTAL\_VEGAMAG column as our preferred magnitude. These filters sample the Rayleigh-Jeans tail of the SED and help anchor the slope of the NIRSpec spectrum.


\subsection{Flux calibration and derivation of fundamental parameters}

We calibrated the spectrum of W0146 to its absolute magnitude considering the parallax from~\citet{2021ApJS..253....7K} and 
calculated fundamental parameters semi-empirically from a spectral energy distribution (SED) built using \texttt{SEDkit}~\citep{2015ApJ...810..158F,2020ascl.soft11014F}. The results are shown in Table~\ref{tab:SEDfit}. SEDkit derives bolometric luminosity by integrating over the NIRSpec spectrum anchored by the 2 MIRI photometric points, and then uses this \lbol and a broad assumption of ages (0.5-10\,Gyr) to estimate radius and mass from evolutionary models. Surface gravity and effective temperature are calculated analytically from the definition of log$g$ and the Stefan-Boltzmann law.
Fundamental parameters for W0146AB were estimated assuming a single overluminous source, rather than a blended-light binary. We used this spectrum to fit both single and binary spectral templates and spectral models to best characterize the components of this system and search for signatures of binarity blended in the unresolved spectrum of a T+Y dwarf binary, as described in the following sections. The JWST data for W0146 presented in this paper can be found in MAST: \dataset[http://dx.doi.org/10.17909/jrmn-sr24]{http://dx.doi.org/10.17909/jrmn-sr24}.

\begin{deluxetable*}{lcc}
\tablecolumns{3}
\tablenum{1}
\tablewidth{0pt}
\setlength{\tabcolsep}{0.05in}
\tablecaption{
Properties of the combined-light system W0146AB.\label{tab:SEDfit}}
\tablehead{\colhead{Parameter} & \colhead{Value} & \colhead{Ref.}}
\startdata
\cutinhead{Astrometric and Kinematic Properties}
R.A. (deg) & 26.735358(2.0)\tablenotemark{a} & \citet{2021ApJS..253....7K}\\
Declination (deg) &  42.569399(1.9)\tablenotemark{a} & \citet{2021ApJS..253....7K} \\
Parallax (mas) & $51.7\pm2.0$ & \citet{2021ApJS..253....7K} \\
Proper motion in R.A. (mas/yr) & {$-451.6\pm0.9$} & \citet{2021ApJS..253....7K} \\
Proper motion in Dec. (mas/yr) & {$-33.1\pm0.9$} & \citet{2021ApJS..253....7K} \\
\cutinhead{Spectrophotometric Properties}
NIR Spectral Type & T9+Y0 & \citet{2021ApJS..253....7K} \\
MKO Y (mag) & 21.60$\pm$0.15 & \citet{2015ApJ...803..102D}\\
MKO J (mag) & 20.69$\pm$0.07 & \citet{2015ApJ...803..102D}\\
MKO H (mag) & 21.30$\pm$0.12 & \citet{2015ApJ...803..102D}\\
IRAC [3.6] (mag) & 17.360$\pm$0.089 & \citet{2018ApJ...867..109M}\\
IRAC [4.5] (mag) & 15.069$\pm$0.022 & \citet{2018ApJ...867..109M} \\ 
\cutinhead{SED-derived fundamental parameters}
$L_\mathrm{bol}$ (erg/s) & $(3.21\pm0.36)\times10^{27}$ & This paper\\
$T_\mathrm{eff}$ (K)\tablenotemark{b} & $571^{+33}_{-19}$ & This paper\\
log $g$ (dex)\tablenotemark{b} & $4.92^{+0.13}_{-0.27}$ & This paper\\
Mass ($M_\mathrm{Jup}$)\tablenotemark{b} & $29^{+6}_{-12}$ & This paper\\
Radius ($R_\mathrm{Jup}$)\tablenotemark{b} & $0.909^{+0.09}_{-0.04}$ & This paper\\
\enddata
\tablenotetext{a}{Uncertainties in parenthesis are in units of mas.}
\tablenotetext{b}{The SED fit assumes an age of $4.5\pm4.0$\,Gyr and uses the unresolved, blended-light, binary spectrum, to estimate fundamental parameters for the source. Therefore, these values are likely unphysical for a binary system.}
\end{deluxetable*}

\section{Spectral Fitting}~\label{sec:splib}

\subsection{Empirical template library}

We used the JWST/
NIRSpec spectra from the GO 2124 program (PI: J. Faherty) to build a homogeneous spectral library of absolute flux-calibrated SEDs to carry out single and binary empirical fits. The 12 targets from this program (including our target, W0146) were selected based on their [3.6]-[4.5] colors and absolute magnitude in [4.5], such that they span median and extreme brightness across a 1.5-mag range of colors and temperatures between 300--700\,K (Faherty et al., in prep.). The publicly available G395M spectrum ($R\sim1000$) of the coldest brown dwarf, WISE J085510.83-071442.5 (hereafter: W0855; \citealt{2024AJ....167....5L}), was also included in our library. All spectra were flux calibrated by the JWST pipeline and scaled to their absolute magnitudes with \texttt{SEDkit} in the same way as W0146 (see Section~\ref{sec:obs}, and more in Faherty et al., in prep.). Existing photometric or spectroscopic literature data on any given source, as well as 3 MIRI photometric points at 10, 15, and $18\,\mu$m to ground the long wavelength portion of the SED, were included in the construction of the SEDs (Faherty et al. in prep).

We interpolated the single templates to the wavelength range and spectral resolution of 
W0146 using a simple spline function while conserving flux to simplify our comparisons. Binary templates were built by adding single templates together. A total of 78 binary templates were built from the 12 single SED (excluding W0146) matching every two as a combination with replacement to also consider pairs with two identical SEDs. 

\subsection{Model library}

We built a spectral model library using the state-of-the-art 1D radiative-convective equilibrium spectral models of~\citet{2023ApJ...950....8L} in their extended version (Lacy et al., in prep.). These models span effective temperatures of 200--800\,K spaced every 25\,K between 250--600\,K and every 50\,K at hotter temperatures, surface gravity between 3.5--5.0\,dex every 0.25\,dex, and metallicity options at $Z/Z_{\odot}$ = 0.316, 1.00, and 3.16 (or equivalently $\mathrm{[Fe/H]} = -0.5, 1.0, +0.5$), and 3 possible ratios of mixing length to pressure scale height H$_{mix}$ = 0.01, 0.10, and 1.00. The models provide a subgrid of three discrete cloud options (clear, vertically compact and vertically extended water clouds, all with disequilibrium chemistry in both CO-CH$_4$ and N$_2$-NH$_3$) between 250--400\,K, log surface gravity between 4.0--5.0, and for metallicities of $Z/Z_{\odot}$ = 0.316 and 1.00, with mixing length ratio H$_{mix}$ = 1. These grids provide a total of 1365 model spectra. We scaled the models' flux density per unit surface area in units of $[f_{\lambda}] = erg\,s^{-1}\,cm^{-2}\,\mathrm{\AA}^{-1}$ to their absolute magnitude by assuming a radius of $1\,R_{\mathrm{Jup}}$, and a distance of 10\,pc, scaling the flux by $(R/d)^2$. All model spectra were interpolated to the wavelength range (down from $0.5-30\,\mu$m to $2.87-5.14\,\mu$m) and spectral resolution (down from $R\sim4340$ to $R\sim2700$) of 
W0146's SED.


From the single model library, we built binary model spectra by combining any two single models, totaling 930,930 binary combinations. We set a few requirements to exclude unphysical models: (1) surface gravity must be log $g \geq 4$, which dropped 436,415 binary models, (2) both primary and secondary models must have the same metallicity for a bona fide binary system, reducing available binary models to 328,375, (3) combinations must have a primary warmer than the secondary, which drops another 77,473 binary models, (4) and surface gravity of the primary must be larger than that of the secondary, dropping another 36,059 binary models, collectively reducing the number of viable binary models to 52,608. 

We also tried an experiment where we constrained the primary and secondary to be coeval by restricting their ages to be within 0.2\,Gyr of each other. We used the evolutionary models of~\citet{2021ApJ...920...85M} to estimate age and mass from model \teff\ and log $g$. We conservatively set a tolerance of 0.2\,Gyr for coevality and excluded models where the surface gravities were different despite having the same \teff. The rationale for this condition is that in a coeval binary system, if both components have the same \teff, then they must have the same mass, radius, and therefore, surface gravity. This coevality requirement reduced the number of available binary models for comparison to 16,674. We discuss results from comparisons to both binary model libraries in Section~\ref{sec:discussion}.



\subsection{Fit ranking with $\chi^2$ goodness-of-fit statistic}\label{sec:chi2}

We compared our target to both single and binary templates and models and assessed the goodness-of-fit with a $\chi^2$ statistic following the prescription of~\citet{2010ApJ...710.1142B} and~\citet{2014ApJ...794..143B}: 

\begin{equation}
\chi^2 = \sum_{\lambda}\frac{(O_{\lambda} - \alpha E_{\lambda})^2}{\sigma_{\lambda}^2}
\end{equation}

\noindent where the subscript $\lambda$ indicates individual wavelength bins across the spectrum, $O_{\lambda}$ is the observed flux of the target at wavelength $\lambda$, $\alpha$ is the factor that minimizes $\chi^2$, $E_\lambda$ is the expected flux at wavelength $\lambda$ from either model or template spectrum, and $\sigma^2_{\lambda}$ is the variance of the target's flux for each wavelength bin (i.e., the square of the flux uncertainty at wavelength $\lambda$). All wavelength bins are weighted equally, except for the $0.10\,\mu$m gap in the spectra between $3.69-3.79\,\mu$m which is weighted zero in target, templates, and models. In order to minimize the $\chi^2$ function, we set the derivative $\frac{d\chi^2}{d\alpha} = 0$ to zero (see~\citealt{2008ApJ...678.1372C,2021ApJ...920...99S}). We also calculated a reduced $\chi^2$ by dividing the chi-squared statistic over the number of degrees of freedom in the comparison, $\chi^2_R = \chi^2/dof$. The number of degrees of freedom is estimated as $dof = N - \nu$ with $N=2675$ available data points in each spectrum and $\nu = 1$, i.e. the $\alpha$ parameter that constrains the fit. Despite the 5 parameters that define each model spectrum, we use the same number of degrees of freedom as for the template comparisons, since the goodness-of-fit is calculated between target and every model.



\section{Results}~\label{sec:results}

\begin{deluxetable*}{cccccccc}
\tabletypesize{\scriptsize}
\tablecolumns{8}
\tablenum{2}
\tablewidth{0pt}
\setlength{\tabcolsep}{0.05in}
\tablecaption{Best fit empirical templates to W0146.\label{tab:specfit}}
\tablehead{\colhead{Comparison} & \colhead{Name} & \colhead{T$_{eff}$ (K)} & \colhead{log\,g (dex)} & \colhead{Mass (M$_{Jup}$)} & \colhead{Radius (R$_{Jup}$)} &  \colhead{$\chi_R^2$} & \colhead{Ref.}}
\startdata
Best fit single & WISE J053516.80-750024.9 & $560_{-15}^{+28}$ & $4.89_{-0.27}^{+0.14}$ & $28_{-11}^{+6}$ & $0.913_{-0.04}^{+0.09}$ & 7.79 & Faherty et al. (in prep.)\\
\hline
\multirow{2}{*}{Best fit binary} & WISEP J075108.79-763449.6 & $484\pm16$ & $4.20\pm0.25$ & $8.40\pm3.14$ & $1.10\pm0.06$ & \multirow{2}{*}{3.41} & Kiman et al. (in prep.)\\
 & WISE J140518.39+553421.3 & $392_{-15}^{+16}$ & $4.69_{-0.23}^{+0.13}$ & $19_{-8}^{+5}$ & $0.96_{-0.03}^{+0.09}$ & & \citet{2024ApJ...973..107B}, Faherty et al. (in prep.)\\
\enddata
\end{deluxetable*}

\begin{deluxetable*}{ccccccc}
\tablecolumns{7}
\tablenum{3}
\tablewidth{0pt}
\setlength{\tabcolsep}{0.05in}
\tablecaption{Best fit models to W0146 from~\citet{2023ApJ...950....8L}.\label{tab:modelfit}}
\tablehead{\colhead{Comparison} & \colhead{T$_{eff}$ (K)} & \colhead{log\,g (dex)} & \colhead{Z/Z$_{\odot}$} & \colhead{Cloud} & \colhead{H$_{mix}$} & \colhead{$\chi_R^2$}}
\startdata
Best fit single & 550 & 4.25 & 0.316 & clear diseq. & 0.01 & 14.81\\
\hline
\multirow{2}{*}{Best fit binary} & 550 & 4.25 & 0.316 & clear diseq. & 0.01 &  \multirow{2}{*}{12.62}\\
 & 325 & 4.25 & 0.316 & clear diseq. & 0.10 & \\
\enddata
\end{deluxetable*}

\subsection{Single fits}

\begin{figure*}
    \centering
    \includegraphics[width=\textwidth, trim= 150 0 150 40, clip]{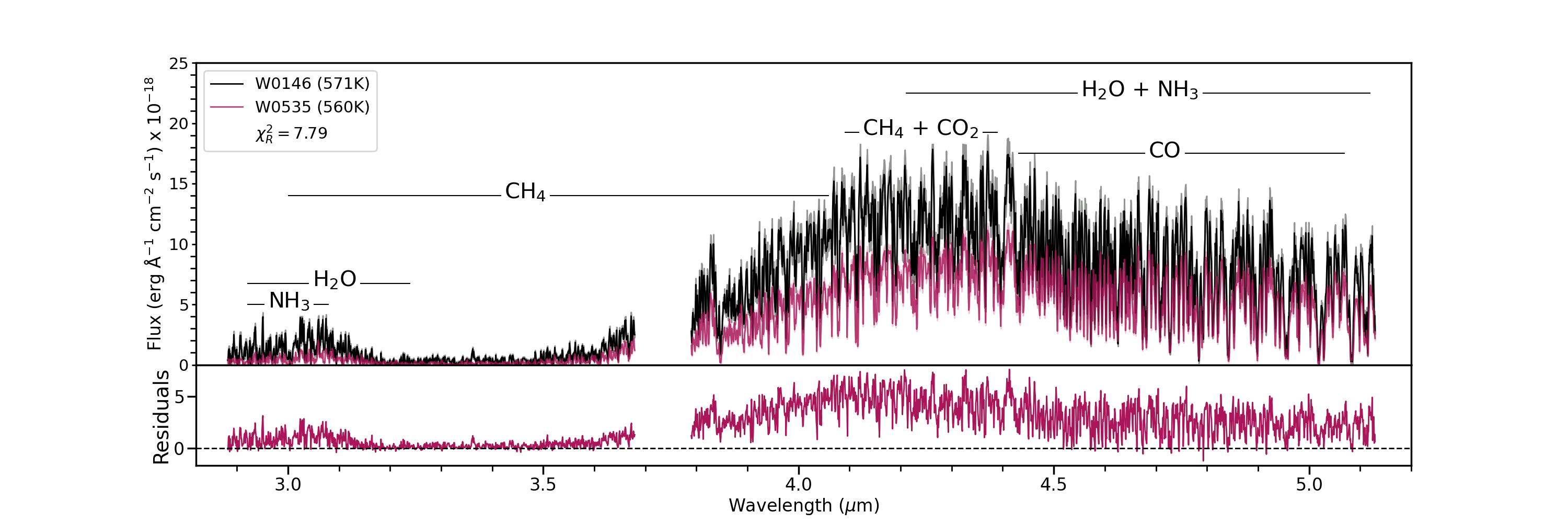}
    \includegraphics[width=\textwidth, trim= 150 0 150 40, clip]{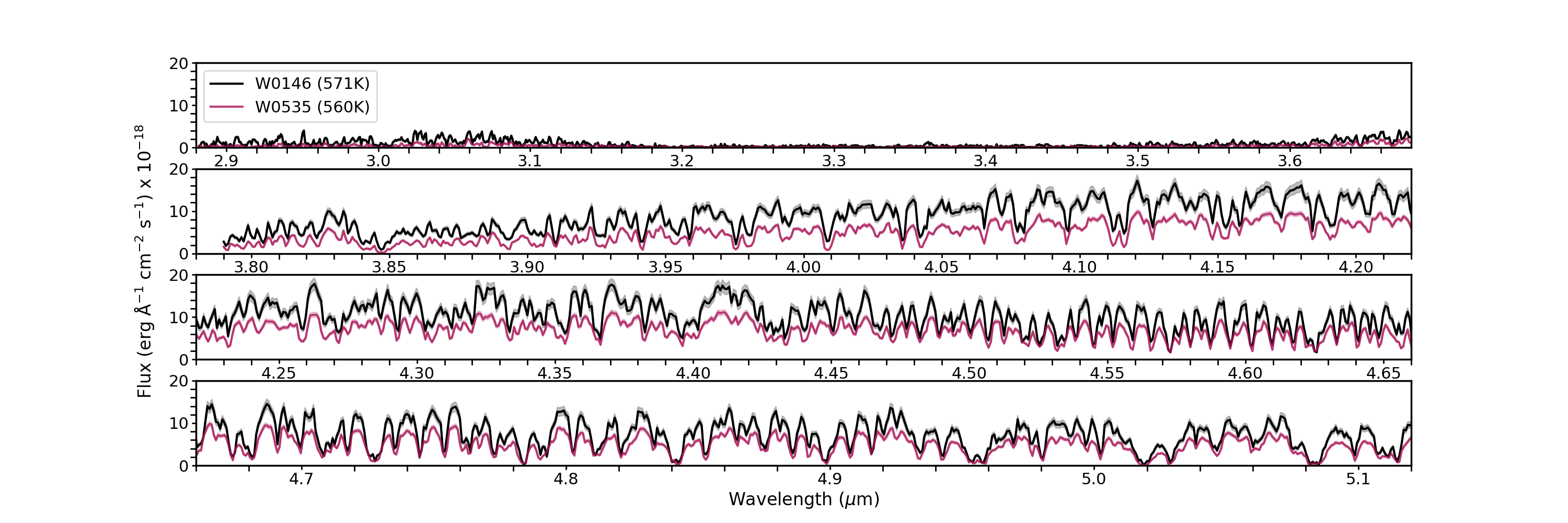}
    \caption{The best fit single template for the blended-light binary W0146 is W0535. \emph{(Top)} Comparison of W0146 (black) and W0535 (dark red) across the full wavelength range, with residuals on the bottom panel resembling the flux added by the secondary to the combined-light spectrum. \emph{(Bottom)} Detailed view of the spectral comparison between these two objects. W0146 is brighter than W0535 at all wavelengths, as expected for a blended-light source compared to a single one of a similar temperature.}
    \label{fig:single_template_fit}
\end{figure*}

\begin{figure*}
    \centering
    \includegraphics[width=\textwidth, trim= 150 0 150 40, clip]{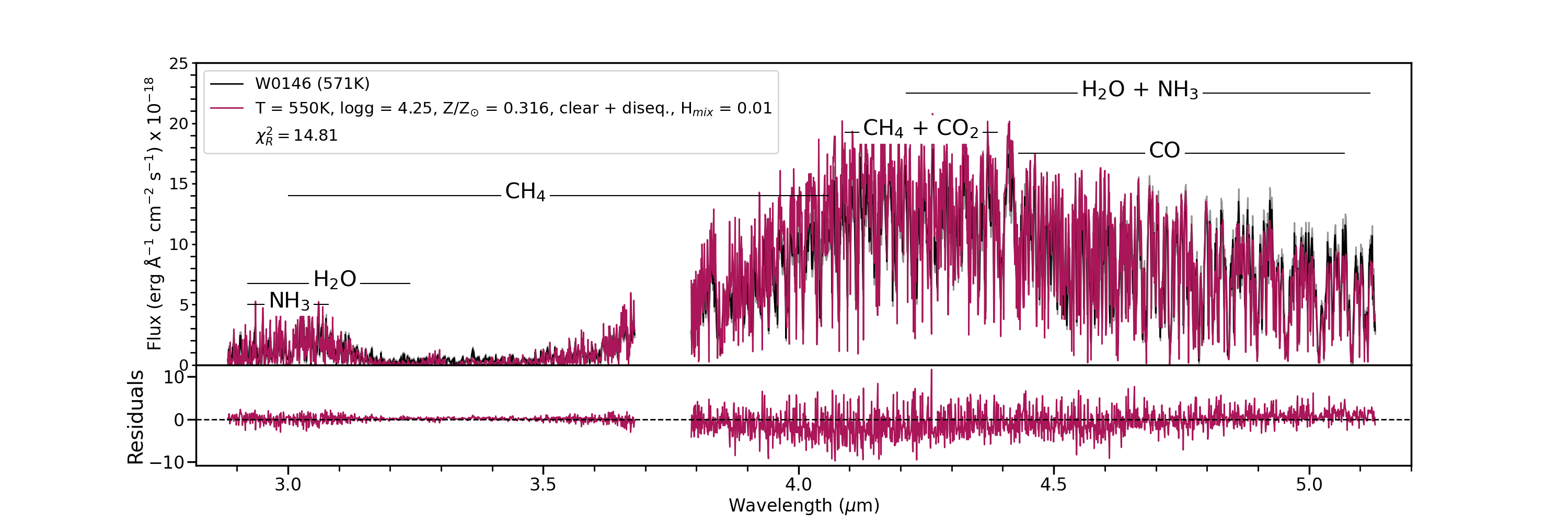}
    \includegraphics[width=\textwidth, trim= 150 0 150 40, clip]{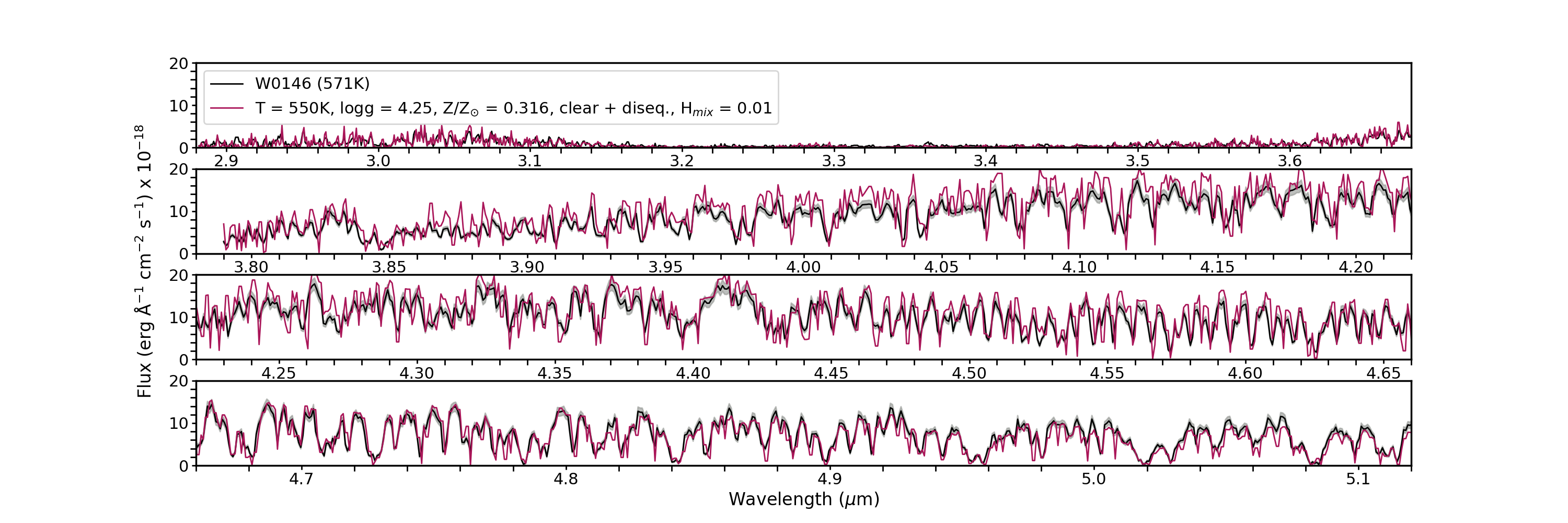}
    \caption{The best fit single model for W0146 is a 550\,K object with a log surface gravity of 4.25, sub-solar metallicity, cloudless atmosphere with disequilibrium chemistry and low vertical mixing ratio. \emph{(Top)} Comparison of W0146 (black) and the best fit model across the full wavelength range, with residuals on the bottom panel showing excess flux from W0146 at wavelengths longer than $4.7\,\mu$m. \emph{(Bottom)} Detailed view of the spectral comparison showing good agreement at shorter wavelengths where the primary is expected to be brighter. The model spectrum was flux calibrated assuming a radius of 1\,$R_{Jup}$ at a distance of 10\,pc.}
    \label{fig:single_model_fit}
\end{figure*}

\subsubsection{Single template fit}\label{sec:single_template_fit}

Even though W0146 is a confirmed binary system, resolved in adaptive optics imaging with Keck~\citep{2015ApJ...803..102D}, we fit it to single templates and single model spectra to understand the difference in spectral features between a blended-light binary spectrum and a single one of comparable temperature (Figure~\ref{fig:single_template_fit}).  It is interesting to note that the best fit single template to our blended-light binary, with a $\chi^2_R = 7.79$, is a potential binary itself: WISE J053516.80-750024.9 (hereafter W0535), an overluminous 496\,K brown dwarf~\citep{2024ApJ...973..107B} suspected to be an unresolved binary~\citep{2021ApJ...918...11L}. Both W0146 and W0535 show a deep, narrow ammonia absorption feature at 3\,$\mu$m~\citep{2023ApJ...951L..48B} and a strong methane absorption feature at 3.84\,$\mu$m. The most striking difference between the two spectra is in the slope of the combined CH$_4$ and CO$_2$ absorption features between 3.85\,$\mu$m and 4.4\,$\mu$m. While still weak, the CO$_2$ absorption feature is slightly more pronounced in W0535, probably due to lower CH$_4$ absorption in W0146 by comparison. Given the similar amount of flux in both sources between 2.9-3.7\,$\mu$m, the difference in slope due to the CH$_4$ feature at 3.85-4.4\,$\mu$m causes W0146 to appear bluer than W0535 in [3.6]-[4.5] Spitzer bands (see Figure~\ref{fig:cmd}). This could be an indication of binarity or simply different carbon disequilibrium chemistry strengths between W0535 and the combined-light spectrum of W0146. The CO absorption band across $4.42-4.95\,\mu$m is similarly weak on both sources. Overall, W0146 is brighter across all wavelengths compared to W0535, as expected for a blended-light binary source of unequal flux components, and shown by the residuals in Figure~\ref{fig:single_template_fit}. However, we cannot rule out the possibility that W0535 is a binary system with cooler components than those of W0146.

\subsubsection{Single model fit}

The best fit single model for W0146 is a 550\,K, slightly metal-poor (Z/Z$_{\odot}$ = 0.316) model with log g = 4.25, disequilibrium chemistry, and a weak mixing length to scale height ratio of H$_{mix}$ = 0.01 implying slow vertical mixing (Figure~\ref{fig:single_model_fit}). 
Overall, the model fits individual features of the W0146 spectral binary with good accuracy (e.g., at CO: 4.5-5.0\,$\mu$m, NH$_3$: 2.9-3.1\,$\mu$m, CH$_4$: 3.0-4.0\,$\mu$m), only varying in flux density with respect to the target. This model has a pronounced ammonia feature at 3\,$\mu$m and methane absorption at 3.84\,$\mu$m both of which distinctly fit the target's features. Disequilibrium chemistry in this best fit model suggests the presence of CO and CO$_2$ in the spectrum of W0146, although the low vertical mixing ratio implies a weak mixing strength which is consistent with the target's weak CO absorption feature between 4.5-5.0\,$\mu$m. 
Such a close agreement between the single model and the target could indicate that the binary system is composed of two objects with similar spectral morphology in this particular wavelength range, or that the secondary is significantly fainter than the primary and therefore does not contribute strongly to the overall flux. While both target and model had been scaled to 1\,\rjup, the scale factor for this combination is 0.78, which means that the flux from the target has been scaled down to 78\% of its value to match the single model's flux. 


\begin{figure*}
    \centering
    \includegraphics[width=\textwidth, trim= 150 0 150 40, clip]{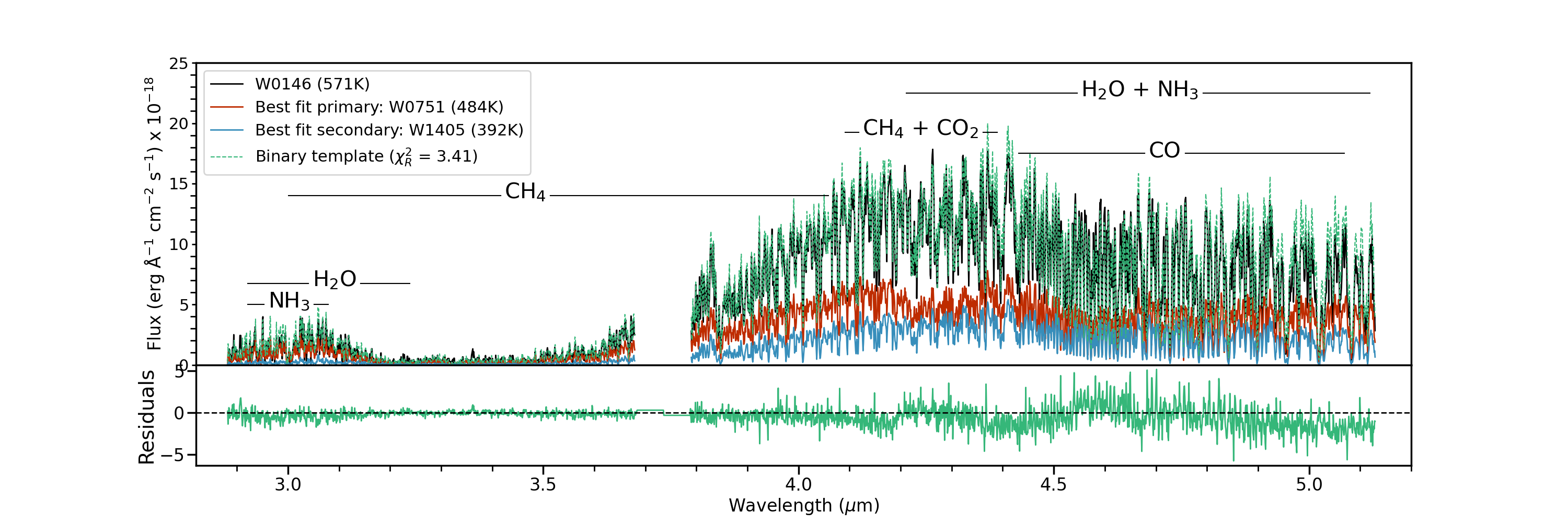}
    \includegraphics[width=\textwidth, trim= 150 0 150 40, clip]{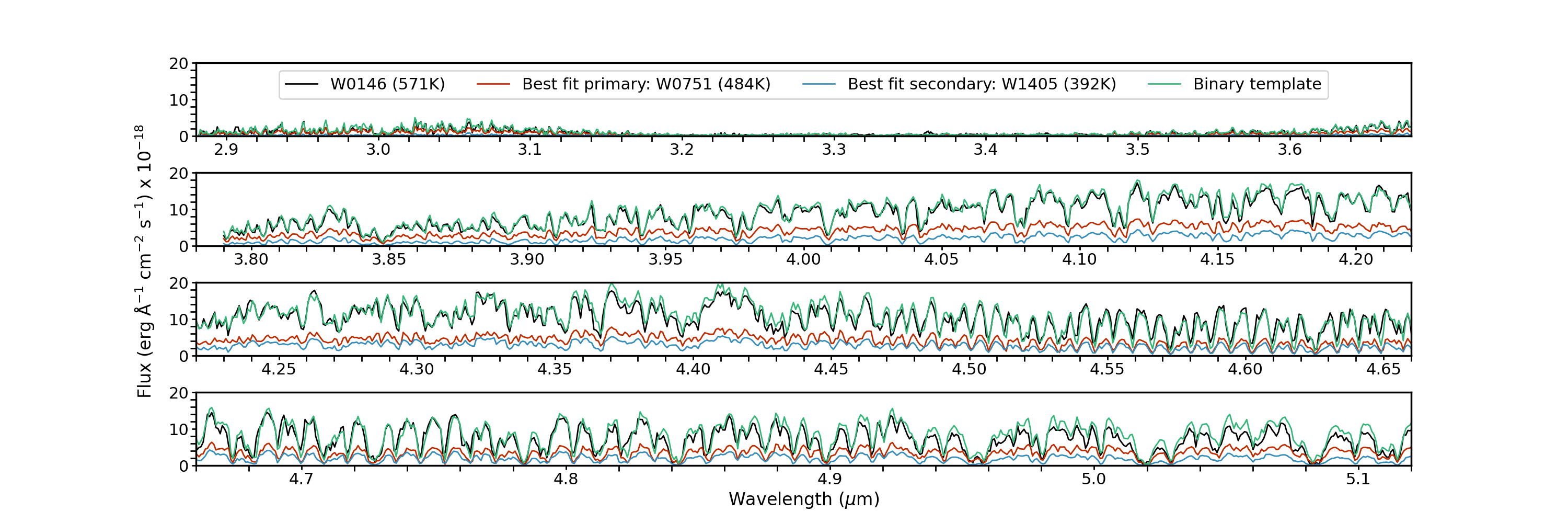}
    \caption{The best fit binary template for W0146 is a combination of W0751 (484\,K) and W1405 (392\,K).}
    \label{fig:binary_template_fit}
\end{figure*}

\begin{figure*}
    \centering
    \includegraphics[width=\textwidth, trim= 150 0 150 0, clip]{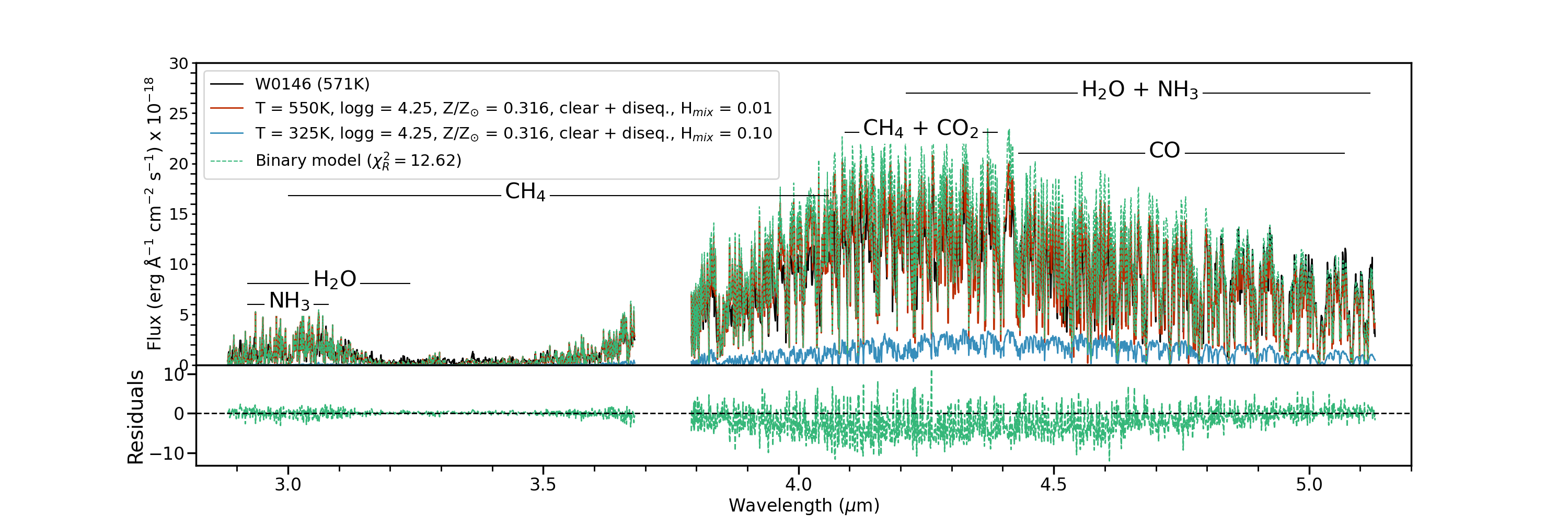}
    \includegraphics[width=\textwidth, trim= 150 0 150 0, clip]{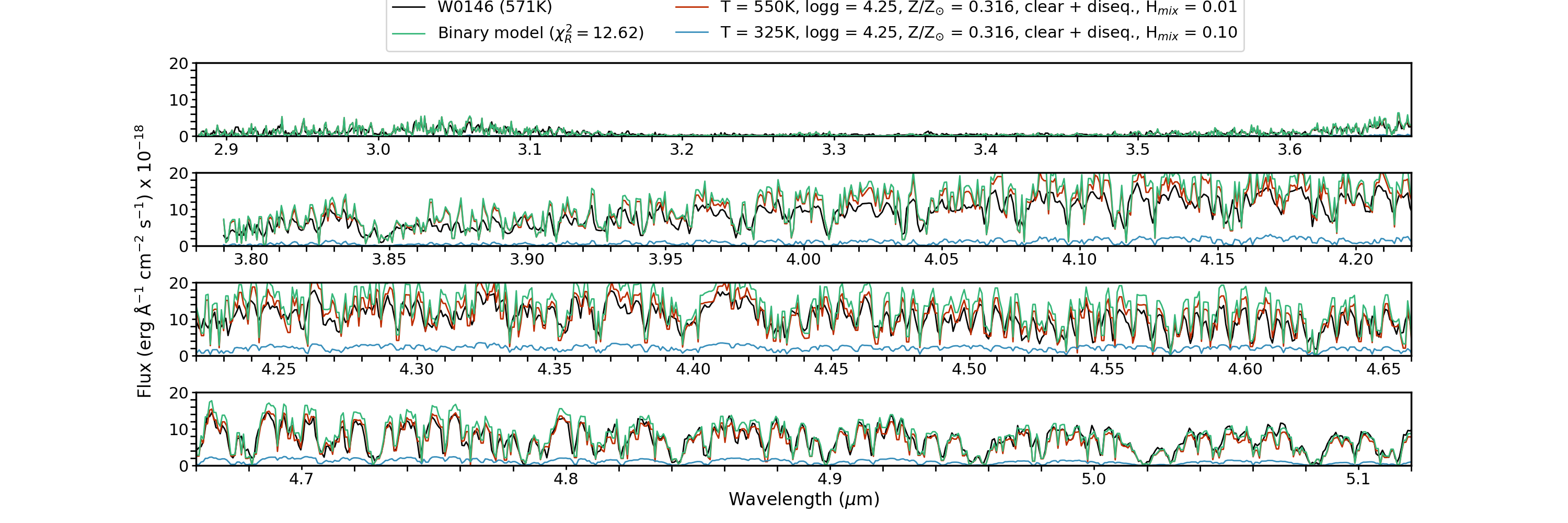}
    \caption{Best fit binary model for W0146.}
    \label{fig:binary_model_fit}
\end{figure*}

\subsection{Binary fits}

\subsubsection{Empirical binary template fit}\label{sec:binary_template_fit}

Based on our library of 12 comparison templates, we find that a combination of the objects WISEP J075108.79-763449.6 (hereafter: W0751) and W1405 yields the lowest $\chi^2_R$ when compared to W0146. This binary template is a significantly better fit to W0146 than the single template, evidenced by the improvement in reduced $\chi^2$ (see Table~\ref{tab:specfit}). W0751, also known as COCONUTS-2b~\citep{2011ApJS..197...19K,2021ApJ...916L..11Z}, is a wide T9 dwarf companion to the young M3 dwarf L 34-26, with an estimated age of 150–800\,Myr and with components separated by 6471\,au. Its published mass from evolutionary models is M = ${6.3}_{-1.9}^{+1.5}$\,M$_{Jup}$~\citep{2021ApJ...916L..11Z}
(see Table~\ref{tab:SEDfit}). W1405 is a variable Y0pec dwarf, with a 3.5\% variability amplitude in [4.5]~\citep{2016ApJ...823..152C}. Its peculiar qualifier arises from a small shift in the H-band compared to the T9 standard, U0722~\citep{2011ApJ...743...50C}. 

Figure~\ref{fig:binary_template_fit} shows the best binary template fit and its components. The 
NIRSpec spectrum of W0751 shows a deep ammonia absorption feature at 3\,$\mu$m, a deep CH$_4$ absorption feature at $3.0-4.0\,\mu$m, as well as a CO$_2$ feature at $4.23-4.42\,\mu$m mixed with CH$_4$, and a wide CO absorption band at $4.5-5.0\,\mu$m. The CO absorption band has a k at $4.65-4.70\,\mu$m. From its G395H spectrum and MIRI photometry, we estimate that this object has a temperature of 484\,K from its SED fit (Faherty et al., in prep.). W1405 on the other hand has a temperature of 435\,K from our SED fits, although a more extensive spectral coverage including MIRI spectroscopy from~\citep{2024ApJ...973..107B} yields 392\,K. Its spectrum has a significantly lower ``continuum'' emission between $2.88-3.12\,\mu$m compared to W0751, which makes its NH$_3$ absorption feature at $3\,\mu$m appear much shallower in comparison. Similarly, the $3.3\,\mu$m methane feature is no longer present, and that wavelength region is rather completely dominated by a mix of water and methane absorption. The slope in the continuum between $3.85-4.18\,\mu$m is steeper in W0751 than W1405, suggesting stronger CH$_4$ absorption in W1405, and making the CO$_2$ absorption feature at $4.23-4.42\,\mu$m slightly less noticeable. In summary, the warmer primary, W0751, has several identifying features, including the ammonia feature at $3\,\mu$m, and a broad CO$_2$ absorption feature at $4.23-4.42\,\mu$m. The slightly cooler secondary, W1405, has all of these features significantly diminished, except for the methane dip at $3.3\,\mu$m which is essentially absent.

This combination of features from W0751 and W1405 yields a spectrum that closely resembles that of W0146: (1) clear ammonia and methane absorption features at $3\,\mu$m and 3.84\,$\mu$m, respectively, and (2) a broad, asymmetric CO$_2$ absorption feature at $4.23-4.42\,\mu$m. The asymmetry on the CO$_2$ band appears by the addition of the different slopes from the continuum between $3.85-4.18\,\mu$m, such that there is a peak in the continuum at $4.18\,\mu$m, a minimum at $4.23\,\mu$m, and another peak at $4.41\,\mu$m. The most important contribution from the secondary seems to be, primarily, excess flux between $4-5\,\mu$m, which would shift the [3.6] - [4.5] color of the blended-light spectrum to the red with respect to the color of the primary by itself (see Figure~\ref{fig:cmd}). Therefore, the most obvious hint at a potential blended-light binary might be the [3.6] - [4.5] redder color than typical for its temperature. After all, W1405 and W0751 have similar absolute magnitudes in [4.5] from their SEDs, yet radically different [3.6]-[4.5] colors with different manifestations of their carbon chemistry across CH$_4$, CO, and CO$_2$ features. However, it is important to note that W0751 and W1405 have different ages and surface gravity, inferred by the 150-800\,Myr primary for W0751, and field-age presumption for W1405, as well as unknown metallicity. For a true binary template fit where these properties are shared among components, we need a more extensive template library. 




\subsubsection{Model binary fit}\label{sec:binary_model_fit}

From our library of binary models from~\citet{2023ApJ...950....8L}, we find that the best binary model fit to W0146 is composed of two objects with temperatures of 550\,K and 325\,K, log surface gravity of 4.25, sub-solar metallicities (Z/Z$_{\odot}$ = 0.316) and a cloudless atmosphere with disequilibrium chemistry (Figure~\ref{fig:binary_model_fit}). The binary model is brighter than the target between 4-4.5\,$\mu$m, suggesting that the binary model has less CO$_2$ absorption than the target. In this particular combination of models, the primary has weak vertical mixing, while the secondary has stronger mixing. Together, these two models do not amount to the CO$_2$ absorption seen in W0146. However, the binary model is a close fit to the target between $2.8\,\mu$m and $3.6\,\mu$m, and then again throughout most of the CO absorption feature between 4.65-5.20\,$\mu$m. It is worth noticing that the primary in this best binary fit model combination is the same model ranked highest for the best single fit. This would indicate that either the secondary does not add a significant amount of flux to the primary and the two sources have a large flux ratio, or that the single model by itself contains averaged characteristics from both components of the binary system, particularly as shallower absorption features. 

While the temperatures of these two models do not closely resemble those of the components of the best fit binary template (see Tables~\ref{tab:modelfit} and~\ref{tab:specfit}), the multiple parameters being explored can lead to degeneracies in the best binary model combinations. The best fit binary template combines two objects with \teff = 484\,K and 392\,K, whereas the best fit binary model integrates two 550\,K and 325\,K models. An important limitation in our comparison is the age and surface gravity difference between the two components from the best template fit, exacerbated by our limited knowledge on their individual metallicities and strength of vertical mixing. Figure~\ref{fig:heatmap} shows the most probable combinations of primary and secondary model temperatures and surface gravity, as defined by their reduced $\chi^2_R$. Primary models with effective temperatures between 450-575\,K combined with secondary models with \teff = 250-400\,K yield relatively similar $\chi^2_R$, whereas a log g of 4.75 is strongly preferred. 

\begin{figure*}
    \centering
    \includegraphics[width=\textwidth, trim=50 50 80 30, clip]{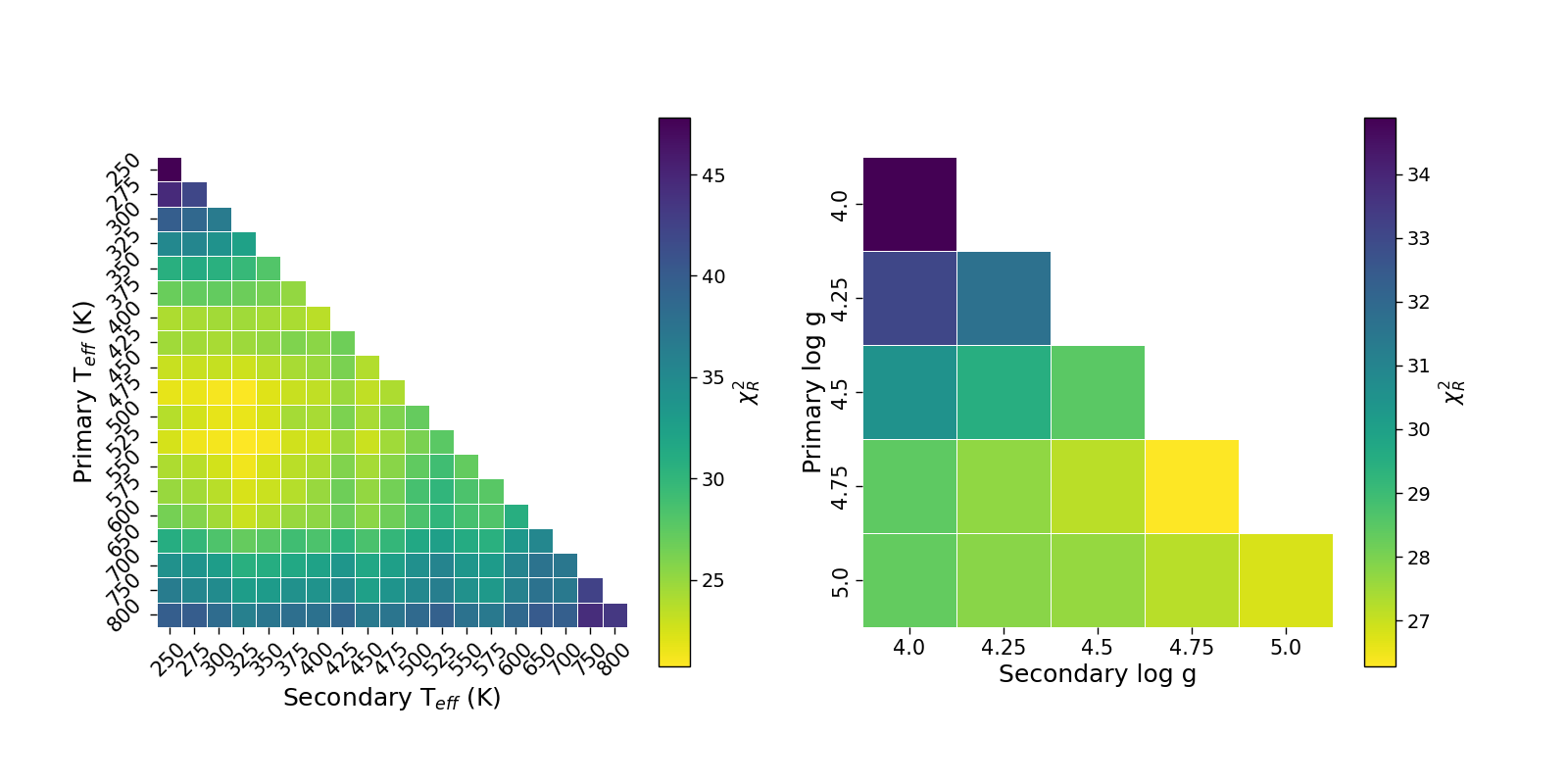}
   \caption{Combinations of models as a function of $\chi^2_R$ for \teff~(\textit{Left}) and surface gravity (\textit{Right}).}
    \label{fig:heatmap}
\end{figure*}


\section{Discussion}\label{sec:discussion}

\subsection{Carbon disequilibrium chemistry hinders cold brown dwarf binary signatures}

From our binary template fits using JWST/
NIRSpec data, we find that the best-fit spectral binary template in the $2.8-5.1\,\mu$m region is a combination of W0751 and W1045. 
In~\citet{2015ApJ...803..102D}, the authors find that a combination of the T9 U0722 and the same W1405 source most closely fit the width of the J and H bands of W0146, in an otherwise mostly featureless near-IR spectrum. W0751 was first discovered in WISE imaging~\citep{2011ApJS..197...19K} and later identified as a comoving companion to a young M3 dwarf~\citep{2021ApJ...916L..11Z}, 8 years after the discovery and near-IR spectral decomposition of W0146~\citep{2015ApJ...803..102D}. The location of W0751 in a [3.6]-[4.5] vs. M$_{[4.5]}$ color-magnitude diagram (CMD) is $0.42\pm0.15$\,mag fainter in absolute magnitude and $0.28\pm0.05$\,mag bluer than U0722, sufficiently close that it serves as additional validation for our spectral template comparison (Figure~\ref{fig:cmd}). A JWST/
NIRSpec spectrum of U0722 is needed to check whether the best fit combination in the near-IR~\citep{2015ApJ...803..102D} continues to be the best binary fit in the $3-5\,\mu$m range.

Based on our binary template analysis, W0146 likely contains a blue primary and a red secondary such that the combination of colors turns the combined-light spectrum of W0146 into a bluer source than its best single fit, W0535. However, while the differences in color are apparent in [3.6]-[4.5] photometry, they are much harder to discern in $2.8-5.1\,\mu$m spectroscopy. Rather than a clear signature of blended-light binarity, the 392\,K secondary has similar yet shallower features compared to the 484\,K primary, such that a blend of both spectra does not produce peculiar individual features. 

The wavelength range where the target and binary template are most dissimilar is between $4.10-5.12\,\mu$m. The range between $3.8-4.2\,\mu$m is primarily affected by CH$_4$ absorption, which is more pronounced in the colder W1405 compared to W0751, and evidenced as increasing flux with wavelength but at different slopes. Together, this sum of fluxes closely approximates the blended-light binary. However, at wavelengths longer than $4.2\,\mu$m, the methane feature starts to mix with CO$_2$ due to disequilibrium chemistry. At the cold effective temperatures of Y dwarfs, the dominant carbon species at equilibrium should be methane; however, strong vertical mixing in these objects dredges CO and CO$_2$ upwards from deeper, hotter atmospheric layers in a timescale shorter than the reaction timescale~\citep[e.g.,][]{2022ApJ...938..107M}, causing a measurable amount of CO and CO$_2$ in their spectra. A recent study by \citet{2024ApJ...973..107B} identified a Y dwarf sequence from JWST NIRSpec and MIRI LRS spectra only at wavelengths shorter than $2\,\mu$m and longer than $8\,\mu$m. The intermediate region centered at $5\mu$m appears to be most sensitive to disequilibrium chemistry and metallicity. Based on our binary template fits, the $4.2-5.0\,\mu$m wavelength range seems particularly sensitive to disequilibrium chemistry (Figure~\ref{fig:binary_template_fit}), where we see absorption by CO and CO$_2$ in both target and binary template, especially when we take into account that the metallicity of both components of a binary system should be the same. However, the residuals in this wavelength region indicate a mismatch between these spectra. The strength of vertical mixing causing the disequilibrium chemistry in the two specific components of our binary template do not reproduce the blended-light spectrum of W0146 well in between $4.2-5.0\,\mu$m. When we compare the binary model fits, the primary model has a weaker vertical mixing than the secondary model, but also very different component \teff~compared to the binary template fit. Since the CO$_2$ absorption feature is stronger in W0751 than in W1405, it is possible that this object has stronger vertical mixing, in turn making its [3.6] - [4.5] color bluer than typical for its temperature and driving the blue color of the combined-light spectrum comparison with W0146.



\subsection{The true diversity of cold worlds}


While the T/Y transition is defined in the near-IR by the appearance of ammonia at 1.53-1.58\,$\mu$m despite its overlap with water and methane~\citep{2011ApJ...743...50C,2002Icar..155..393L,2003ApJ...596..587B,1999ApJ...512..843B}, signs of this transition can also be seen in the 3-5\,$\mu$m regime covered by Spitzer and JWST/NIRSpec. W0146AB is likely composed of two objects with atmospheres similar to those of W0751 and W1405. These two components are classified as T9.5 and Y0pec, respectively, based on their NIR spectral morphology. Their J-K colors are almost identical ($-0.69\pm0.206$ and $-0.55\pm0.134$, respectively;~\citealt{2011ApJS..197...19K,2015ApJ...799...37L,2013ApJ...763..130L,2012ApJ...748...74L}), and their \teff~differs by only $92\pm23$\,K. Therefore, it makes sense that their spectral classification is essentially the same. However, when we look at the 3-5\,$\mu$m range, their absolute magnitudes in [4.5] differ by $0.48\pm0.065$\,mag, and their [3.6]-[4.5] color by $1.02\pm0.155$\,mag.


Figure~\ref{fig:cmd} shows the color-magnitude diagram in Spitzer colors of the late-T and Y dwarf population for objects with parallax errors smaller than 10\% compiled from~\citet{2024ApJS..271...55K}. In the absence of a spectral type classification system in the $3-5\,\mu$m regime, we estimate \teff~from [3.6]-[4.5] color based on the empirical relation of~\citet{2019ApJS..240...19K}. Highlighted sources include W0146 shown as a star, its best fit single template, W0535, shown as a triangle, and its best fit primary and secondary templates, W0751 and W1405, shown as circles. Confirmed T/Y binary systems are shown as squares. 
In addition to a clear correlation between fainter absolute magnitude in [4.5] and redder [3.6]-[4.5] color, surface gravity, metallicity, binarity, and carbon disequilibrium chemistry have a visible effect on the color spread of this population. The best-fit components W0751 and W1405 lie on opposite sides of the population, apparently flanking the T/Y transition in the $3-5\,\mu$m regime; both being presumably single sources of unknown metallicity and different surface gravity owing to the young age of W0751. At these temperatures and wavelengths, disequilibrium chemistry becomes an additional, essential factor changing the shape of the emerging spectra and the overall flux at [3.6] and [4.5] bands~\citep{2020AJ....160...63M,2022ApJ...938..107M}.




\begin{figure*}
    \centering
    \includegraphics[width=\textwidth]{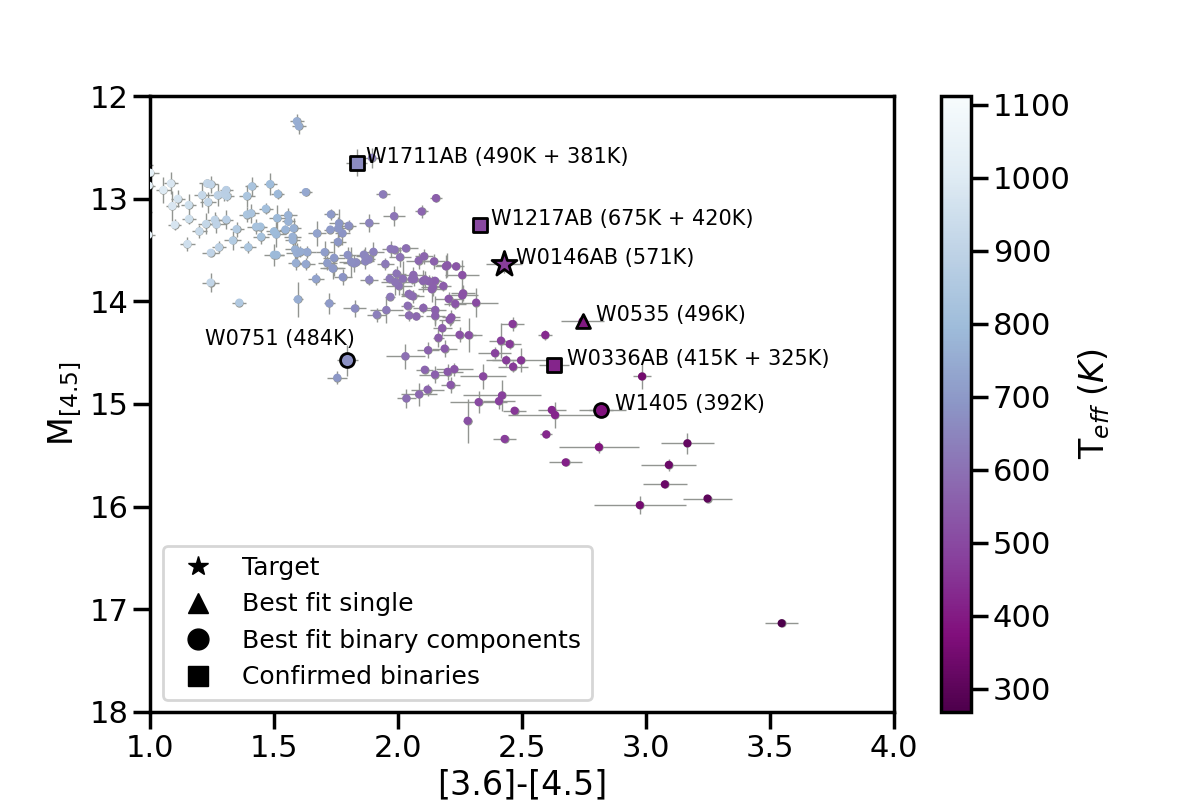}
    \caption{Color-magnitude diagram of late-T and Y dwarfs in Spitzer colors. Effective temperatures have been estimated using color-temperature relations from~\citet{2019ApJS..240...19K}. Our target, W0146, is shown as a 
    star. Its best fit single is W0535, slightly fainter and redder. The components of its best fit binary template are W0751 as a primary, and W1405 as a secondary. The addition of these two spectra would produce a blended-light template that is redder than expected for its temperature. Three other confirmed binary systems straddling the T/Y transition are shown 
    as squares: WISE J1711+3500~\citep{2012ApJ...758...57L}, WISE J1217+1626~\citep{2012ApJ...758...57L}, and W0336AB~\citep{2023ApJ...947L..30C}.}
    \label{fig:cmd}
\end{figure*}

While the \teff~continuously decreases with redder [3.6]-[4.5] colors and fainter absolute magnitudes in [4.5] according to the \citet{2019ApJS..240...19K} relation, there appears to be a slight change in slope around [3.6]-[4.5]$\sim$2.2\,mag. This color corresponds to a \teff~of about 540\,K~\citep{2019ApJS..240...19K}, which is close to the nominal start to the Y dwarf class ($\sim500$\,K) from the near-IR, and to the \teff~at which the Sonora Bobcat models predict a minimum in the eddy diffusion coefficient ($K_{zz}$, the model parameter representing the strength of large-scale vertical mixing) over the late-T/Y dwarf temperature range~(100-700\,K;~\citealt{2020AJ....160...63M}). A low value of $K_{zz}$ indicates a long mixing timescale within an atmosphere, hence a weak vertical mixing. At temperatures lower than $\sim540\,$K, models predict an increase in the $K_{zz}$, thus shortening the mixing timescale and allowing chemical disequilibrium to dominate. Therefore, this change in slope in the CMD at [3.6]-[4.5] = 2.2\,mag which roughly corresponds to the temperature at which the T/Y transition occurs, could also be representative of the temperature at which the strength of carbon disequilibrium chemistry changes in the coldest brown dwarfs. The T/Y transition could then be potentially driven by a change in the strength of the carbon disequilibrium chemistry, mirroring the atmospheric changes that drive the L/T transition at warmer temperatures as an abrupt change in color over a small temperature range~\citep{2002ApJ...571L.151B,2012ApJS..201...19D,2006ApJS..166..585B,2013AN....334...32B}.




\subsection{Multiplicity as a signpost for the formation mechanism of the coldest companions}

Multiplicity fractions are characteristic of stellar and substellar populations. Based on their measured binary fraction, 
we expect that at most $\sim8\%$ of Y dwarfs have equal or lower mass, chemically-identical companions~\citep{2018MNRAS.479.2702F} which formed simultaneously with the primary from the same molecular cloud, even at planetary masses, expected for the majority of Y-dwarfs~\citep{2019ApJ...882L..29M}.  
With only one Y+Y binary system currently known~\citep{2023ApJ...947L..30C}, it is challenging to robustly constrain the multiplicity fraction and binary properties of the population. However, a departure in the decreasing trend of multiplicity fraction, average separation, and power law index of the mass ratio distribution with mass, could signify a change in the formation mechanism of the companions. Therefore, characterizing the orbital properties and occurrence rates of the lowest-mass binaries is essential to understand their formation channels. This effort relies on expanding our current sample of Y dwarfs. Ongoing and planned efforts to search for Y dwarfs with NEO Surveyor~\citep{2023PSJ.....4..224M}, EUCLID~\citep{2024arXiv240513491E} and the Backyard Worlds: Planet 9 citizen science collaboration are key to grow our current list of $\sim50$ confirmed Y dwarfs.


Identifying blended-light binaries is key to probe the multiplicity properties of the Y dwarf population, especially as most Y dwarf binary systems are expected to be tight, with separations in the order of 1\,au~\citep{2018MNRAS.479.2702F}. We have characterized a handful of late-T/Y dwarf binaries to date (WISE J1711+3500~\citep{2012ApJ...758...57L}, WISE J1217+1626~\citep{2012ApJ...758...57L}, and W0336AB~\citep{2023ApJ...947L..30C}), and are now adding W0146 (see also~\citealt{2015ApJ...803..102D}) as another crucial low-mass binary. Several more sources merit a closer investigation to explore their nature as unresolved binary systems based on their overluminous location in the CMD (Figure~\ref{fig:cmd}). These short orbital periods make them amenable for astrometric follow-up within a few JWST cycles, linking spectral type to dynamical mass. 


\section{Conclusions}~\label{sec:conclusions}

In this paper, we present a spectral binary analysis for the unresolved JWST/NIRSpec medium resolution spectrum of the brown dwarf binary W0146. This source has been characterized in the past as a combination of two unusually faint T9 and Y0 dwarfs~\citep{2015ApJ...803..102D}. A subsequent parallax measurement places the object twice as far as previously measured, such that the components no longer need to have unusual brightness. With this new parallax, we update the projected separation between components to $1.69\pm0.106$\,au. This tightly-separated binary is unresolved in JWST/
NIRSpec spectroscopy, thus becoming the first brown dwarf spectral binary characterized with JWST. Future multi-band differential photometry with JWST/
NIRCam 
will be helpful to constrain the flux differences across wavelength and estimate a mass ratio.

Using a library of JWST/NIRSpec spectra, we find that the combination of W0751 (484\,K) and W1405 (392\,K) best fits the spectrum of W0146. A binary model built from the~\citet{2023ApJ...950....8L} self-consistent model grid, yields components with a different set of primary and secondary temperatures (550\,K and 325\,K). By examining the component spectra, we notice that the spectral differences between the primary and secondary are minor, with the warmer primary showing deeper absorption features of ammonia, methane, and carbon dioxide. However, while the difference in temperature between the two components in the binary template fit is small ($92\pm23$\,K), they have a large color difference in [3.6]-[4.5] ($1.02\pm0.155$\,mag), which could be suggestive of an important change in atmospheric processes, akin to the L/T transition at warmer temperatures yet more subtle, such as the dominance of disequilibrium chemistry.

This source has been monitored through astrometry for a decade and should yield an orbit and dynamical masses within the next few years~\citep{2015ApJ...803..102D}. Continued detailed characterization of this binary source is crucial to benchmark this system and understand its physical parameters and spectral features anchoring their age and composition on each other. Growing the library of JWST/NIRSpec spectra at these extremely cold temperatures and mid-IR wavelengths is key to disentangle the complex effects of disequilibrium chemistry, variability and blended-light binarity in the coldest brown dwarfs.


\begin{acknowledgments}
J.F. acknowledges the Heising-Simons Foundation, the National Science Foundation (Award Nos. 2009177 and 1909776) and NASA (Award No. 80NSSC22K0142). B.L. acknowledges support from the Heising-Simons Foundation via the 51 Pegasi b Fellowship. B.B. acknowledges support from the UK Research and Innovation Science and Technology Facilities Council (Grant No. ST\/X001091\/1). J.M.V. acknowledges support from a university research fellowship funded by the Royal Society and Science Foundation Ireland (URF\/1\/221932). This work was performed in part using high-performance computing equipment at Amherst College obtained under National Science Foundation Grant No. 2117377. Some of the data presented in this paper were obtained from the Mikulski Archive for Space Telescopes (MAST) at the Space Telescope Science Institute. The specific observations analyzed can be accessed via \dataset[http://dx.doi.org/10.17909/jrmn-sr24]{http://dx.doi.org/10.17909/jrmn-sr24}. STScI is operated by the Association of Universities for Research in Astronomy, Inc., under NASA contract NAS5–26555. Support to MAST for these data is provided by the NASA Office of Space Science via grant NAG5–7584 and by other grants and contracts.

\end{acknowledgments}

\vspace{5mm}
\facilities{JWST(NIRSpec)}

\software{SPLAT \citep{2017ASInC..14....7B}}

\bibliography{BDlibrary_AandA}{}

\begin{thebibliography}{}
\expandafter\ifx\csname natexlab\endcsname\relax\def\natexlab#1{#1}\fi
\providecommand{\url}[1]{\href{#1}{#1}}
\providecommand{\dodoi}[1]{doi:~\href{http://doi.org/#1}{\nolinkurl{#1}}}
\providecommand{\doeprint}[1]{\href{http://ascl.net/#1}{\nolinkurl{http://ascl.net/#1}}}
\providecommand{\doarXiv}[1]{\href{https://arxiv.org/abs/#1}{\nolinkurl{https://arxiv.org/abs/#1}}}

\bibitem[{{Allen}(2007)}]{2007ApJ...668..492A}
{Allen}, P.~R. 2007, \apj, 668, 492, \dodoi{10.1086/521207}

\bibitem[{{Bardalez Gagliuffi} {et~al.}(2019){Bardalez Gagliuffi},
  {Ward-Duong}, {Faherty}, {Greenbaum}, {Marocco}, {Burgasser}, {Bate},
  {Dupuy}, {Gelino}, {Sahlmann}, {Martinache}, {Meyer}, {Konopacky}, \&
  {Stephens}}]{2019BAAS...51c.285B}
{Bardalez Gagliuffi}, D., {Ward-Duong}, K., {Faherty}, J., {et~al.} 2019,
  \baas, 51, 285.
\newblock \doarXiv{1903.06699}

\bibitem[{{Bardalez Gagliuffi} {et~al.}(2021){Bardalez Gagliuffi}, {Faherty},
  {Li}, {Brandt}, {Williams}, {Brandt}, \& {Gelino}}]{2021ApJ...922L..43B}
{Bardalez Gagliuffi}, D.~C., {Faherty}, J.~K., {Li}, Y., {et~al.} 2021, \apjl,
  922, L43, \dodoi{10.3847/2041-8213/ac382c}

\bibitem[{{Bardalez Gagliuffi} {et~al.}(2014){Bardalez Gagliuffi}, {Burgasser},
  {Gelino}, {Looper}, {Nicholls}, {Schmidt}, {Cruz}, {West}, {Gizis}, \&
  {Metchev}}]{2014ApJ...794..143B}
{Bardalez Gagliuffi}, D.~C., {Burgasser}, A.~J., {Gelino}, C.~R., {et~al.}
  2014, \apj, 794, 143, \dodoi{10.1088/0004-637X/794/2/143}

\bibitem[{{Beichman} {et~al.}(2014){Beichman}, {Gelino}, {Kirkpatrick},
  {Cushing}, {Dodson-Robinson}, {Marley}, {Morley}, \&
  {Wright}}]{2014ApJ...783...68B}
{Beichman}, C., {Gelino}, C.~R., {Kirkpatrick}, J.~D., {et~al.} 2014, \apj,
  783, 68, \dodoi{10.1088/0004-637X/783/2/68}

\bibitem[{{Beiler} {et~al.}(2023){Beiler}, {Cushing}, {Kirkpatrick},
  {Schneider}, {Mukherjee}, \& {Marley}}]{2023ApJ...951L..48B}
{Beiler}, S.~A., {Cushing}, M.~C., {Kirkpatrick}, J.~D., {et~al.} 2023, \apjl,
  951, L48, \dodoi{10.3847/2041-8213/ace32c}

\bibitem[{{Beiler} {et~al.}(2024{\natexlab{a}}){Beiler}, {Cushing},
  {Kirkpatrick}, {Schneider}, {Mukherjee}, {Marley}, {Marocco}, \&
  {Smart}}]{2024arXiv240708518B}
---. 2024{\natexlab{a}}, arXiv e-prints, arXiv:2407.08518,
  \dodoi{10.48550/arXiv.2407.08518}

\bibitem[{{Beiler} {et~al.}(2024{\natexlab{b}}){Beiler}, {Cushing},
  {Kirkpatrick}, {Schneider}, {Mukherjee}, {Marley}, {Marocco}, \&
  {Smart}}]{2024ApJ...973..107B}
---. 2024{\natexlab{b}}, \apj, 973, 107, \dodoi{10.3847/1538-4357/ad6301}

\bibitem[{{Burgasser}(2004)}]{2004ApJS..155..191B}
{Burgasser}, A.~J. 2004, \apjs, 155, 191, \dodoi{10.1086/424386}

\bibitem[{{Burgasser}(2013)}]{2013AN....334...32B}
---. 2013, Astronomische Nachrichten, 334, 32, \dodoi{10.1002/asna.201211763}

\bibitem[{{Burgasser} {et~al.}(2010){Burgasser}, {Cruz}, {Cushing}, {Gelino},
  {Looper}, {Faherty}, {Kirkpatrick}, \& {Reid}}]{2010ApJ...710.1142B}
{Burgasser}, A.~J., {Cruz}, K.~L., {Cushing}, M., {et~al.} 2010, \apj, 710,
  1142, \dodoi{10.1088/0004-637X/710/2/1142}

\bibitem[{{Burgasser} {et~al.}(2006{\natexlab{a}}){Burgasser}, {Geballe},
  {Leggett}, {Kirkpatrick}, \& {Golimowski}}]{2006ApJ...637.1067B}
{Burgasser}, A.~J., {Geballe}, T.~R., {Leggett}, S.~K., {Kirkpatrick}, J.~D.,
  \& {Golimowski}, D.~A. 2006{\natexlab{a}}, \apj, 637, 1067,
  \dodoi{10.1086/498563}

\bibitem[{{Burgasser} {et~al.}(2006{\natexlab{b}}){Burgasser}, {Kirkpatrick},
  {Cruz}, {Reid}, {Leggett}, {Liebert}, {Burrows}, \&
  {Brown}}]{2006ApJS..166..585B}
{Burgasser}, A.~J., {Kirkpatrick}, J.~D., {Cruz}, K.~L., {et~al.}
  2006{\natexlab{b}}, \apjs, 166, 585, \dodoi{10.1086/506327}

\bibitem[{{Burgasser} {et~al.}(2002){Burgasser}, {Marley}, {Ackerman},
  {Saumon}, {Lodders}, {Dahn}, {Harris}, \&
  {Kirkpatrick}}]{2002ApJ...571L.151B}
{Burgasser}, A.~J., {Marley}, M.~S., {Ackerman}, A.~S., {et~al.} 2002, \apjl,
  571, L151, \dodoi{10.1086/341343}

\bibitem[{{Burgasser} {et~al.}(2007){Burgasser}, {Reid}, {Siegler}, {Close},
  {Allen}, {Lowrance}, \& {Gizis}}]{2007prpl.conf..427B}
{Burgasser}, A.~J., {Reid}, I.~N., {Siegler}, N., {et~al.} 2007, Protostars and
  Planets V, 427

\bibitem[{{Burgasser} \& {Splat Development Team}(2017)}]{2017ASInC..14....7B}
{Burgasser}, A.~J., \& {Splat Development Team}. 2017, in Astronomical Society
  of India Conference Series, Vol.~14, Astronomical Society of India Conference
  Series, 7--12.
\newblock \doarXiv{1707.00062}

\bibitem[{{Burrows} \& {Sharp}(1999)}]{1999ApJ...512..843B}
{Burrows}, A., \& {Sharp}, C.~M. 1999, \apj, 512, 843, \dodoi{10.1086/306811}

\bibitem[{{Burrows} {et~al.}(2003){Burrows}, {Sudarsky}, \&
  {Lunine}}]{2003ApJ...596..587B}
{Burrows}, A., {Sudarsky}, D., \& {Lunine}, J.~I. 2003, \apj, 596, 587,
  \dodoi{10.1086/377709}

\bibitem[{{Calissendorff} {et~al.}(2023){Calissendorff}, {De Furio}, {Meyer},
  {Albert}, {Aganze}, {Ali-Dib}, {Bardalez Gagliuffi}, {Baron}, {Beichman},
  {Burgasser}, {Cushing}, {Faherty}, {Fontanive}, {Gelino}, {Gizis},
  {Greenbaum}, {Kirkpatrick}, {Leggett}, {Martinache}, {Mary}, {N'Diaye},
  {Pope}, {Roellig}, {Sahlmann}, {Sivaramakrishnan}, {Thorngren}, {Ygouf}, \&
  {Vandal}}]{2023ApJ...947L..30C}
{Calissendorff}, P., {De Furio}, M., {Meyer}, M., {et~al.} 2023, \apjl, 947,
  L30, \dodoi{10.3847/2041-8213/acc86d}

\bibitem[{{Close} {et~al.}(2003){Close}, {Siegler}, {Freed}, \&
  {Biller}}]{2003ApJ...587..407C}
{Close}, L.~M., {Siegler}, N., {Freed}, M., \& {Biller}, B. 2003, \apj, 587,
  407, \dodoi{10.1086/368177}

\bibitem[{{Cushing} {et~al.}(2008){Cushing}, {Marley}, {Saumon}, {Kelly},
  {Vacca}, {Rayner}, {Freedman}, {Lodders}, \& {Roellig}}]{2008ApJ...678.1372C}
{Cushing}, M.~C., {Marley}, M.~S., {Saumon}, D., {et~al.} 2008, \apj, 678,
  1372, \dodoi{10.1086/526489}

\bibitem[{{Cushing} {et~al.}(2011){Cushing}, {Kirkpatrick}, {Gelino},
  {Griffith}, {Skrutskie}, {Mainzer}, {Marsh}, {Beichman}, {Burgasser},
  {Prato}, {Simcoe}, {Marley}, {Saumon}, {Freedman}, {Eisenhardt}, \&
  {Wright}}]{2011ApJ...743...50C}
{Cushing}, M.~C., {Kirkpatrick}, J.~D., {Gelino}, C.~R., {et~al.} 2011, \apj,
  743, 50, \dodoi{10.1088/0004-637X/743/1/50}

\bibitem[{{Cushing} {et~al.}(2016){Cushing}, {Hardegree-Ullman}, {Trucks},
  {Morley}, {Gizis}, {Marley}, {Fortney}, {Kirkpatrick}, {Gelino}, {Mace}, \&
  {Carey}}]{2016ApJ...823..152C}
{Cushing}, M.~C., {Hardegree-Ullman}, K.~K., {Trucks}, J.~L., {et~al.} 2016,
  \apj, 823, 152, \dodoi{10.3847/0004-637X/823/2/152}

\bibitem[{{Dupuy} \& {Kraus}(2013)}]{2013Sci...341.1492D}
{Dupuy}, T.~J., \& {Kraus}, A.~L. 2013, Science, 341, 1492,
  \dodoi{10.1126/science.1241917}

\bibitem[{{Dupuy} \& {Liu}(2012)}]{2012ApJS..201...19D}
{Dupuy}, T.~J., \& {Liu}, M.~C. 2012, \apjs, 201, 19,
  \dodoi{10.1088/0067-0049/201/2/19}

\bibitem[{{Dupuy} {et~al.}(2015){Dupuy}, {Liu}, \&
  {Leggett}}]{2015ApJ...803..102D}
{Dupuy}, T.~J., {Liu}, M.~C., \& {Leggett}, S.~K. 2015, \apj, 803, 102,
  \dodoi{10.1088/0004-637X/803/2/102}

\bibitem[{{Euclid Collaboration} {et~al.}(2024){Euclid Collaboration},
  {Mellier}, {Abdurro'uf}, {Acevedo Barroso}, {Ach{\'u}carro}, {Adamek},
  {Adam}, {Addison}, {Aghanim}, {Aguena}, {Ajani}, {Akrami}, {Al-Bahlawan},
  {Alavi}, {Albuquerque}, {Alestas}, {Alguero}, {Allaoui}, {Allen}, {Allevato},
  {Alonso-Tetilla}, {Altieri}, {Alvarez-Candal}, {Alvi}, {Amara}, {Amendola},
  {Amiaux}, {Andika}, {Andreon}, {Andrews}, {Angora}, {Angulo}, {Annibali},
  {Anselmi}, {Anselmi}, {Arcari}, {Archidiacono}, {Aric{\`o}}, {Arnaud},
  {Arnouts}, {Asgari}, {Asorey}, {Atayde}, {Atek}, {Atrio-Barandela}, {Aubert},
  {Aubourg}, {Auphan}, {Auricchio}, {Aussel}, {Aussel}, {Avelino},
  {Avgoustidis}, {Avila}, {Awan}, {Azzollini}, {Baccigalupi}, {Bachelet},
  {Bacon}, {Baes}, {Bagley}, {Bahr-Kalus}, {Balaguera-Antolinez}, {Balbinot},
  {Balcells}, {Baldi}, {Baldry}, {Balestra}, {Ballardini}, {Ballester},
  {Balogh}, {Ba{\~n}ados}, {Barbier}, {Bardelli}, {Baron}, {Barreiro},
  {Barrena}, {Barriere}, {Barros}, {Barthelemy}, {Bartolo}, {Basset},
  {Battaglia}, {Battisti}, {Baugh}, {Baumont}, {Bazzanini}, {Beaulieu},
  {Beckmann}, {Belikov}, {Bel}, {Bellagamba}, {Bella}, {Bellini}, {Benabed},
  {Bender}, {Benevento}, {Bennett}, {Benson}, {Bergamini}, {Bermejo-Climent},
  {Bernardeau}, {Bertacca}, {Berthe}, {Berthier}, {Bethermin}, {Beutler},
  {Bevillon}, {Bhargava}, {Bhatawdekar}, {Bianchi}, {Bisigello}, {Biviano},
  {Blake}, {Blanchard}, {Blazek}, {Blot}, {Bosco}, {Bodendorf}, {Boenke},
  {B{\"o}hringer}, {Boldrini}, {Bolzonella}, {Bonchi}, {Bonici}, {Bonino},
  {Bonino}, {Bonvin}, {Bon}, {Booth}, {Borgani}, {Borlaff}, {Borsato}, {Bosco},
  {Bose}, {Botticella}, {Boucaud}, {Bouche}, {Boucher}, {Boutigny}, {Bouvard},
  {Bouwens}, {Bouy}, {Bowler}, {Bozza}, {Bozzo}, {Branchini}, {Brando},
  {Brau-Nogue}, {Brekke}, {Bremer}, {Brescia}, {Breton}, {Brinchmann},
  {Brinckmann}, {Brockley-Blatt}, {Brodwin}, {Brouard}, {Brown}, {Bruton},
  {Bucko}, {Buddelmeijer}, {Buenadicha}, {Buitrago}, {Burger}, {Burigana},
  {Busillo}, {Busonero}, {Cabanac}, {Cabayol-Garcia}, {Cagliari}, {Caillat},
  {Caillat}, {Calabrese}, {Calabro}, {Calderone}, {Calura}, {Camacho Quevedo},
  {Camera}, {Campos}, {Canas-Herrera}, {Candini}, {Cantiello}, {Capobianco},
  {Cappellaro}, {Cappelluti}, {Cappi}, {Caputi}, {Cara}, {Carbone}, {Cardone},
  {Carella}, {Carlberg}, {Carle}, {Carminati}, {Caro}, {Carrasco}, {Carretero},
  {Carrilho}, \& {Carron Duque}}]{2024arXiv240513491E}
{Euclid Collaboration}, {Mellier}, Y., {Abdurro'uf}, {et~al.} 2024, arXiv
  e-prints, arXiv:2405.13491, \dodoi{10.48550/arXiv.2405.13491}

\bibitem[{{Faherty} {et~al.}(2024){Faherty}, {Burningham}, {Gagne}, {Suarez},
  {Vos}, {Alejandro Merchan}, {Morley}, {Rowland}, {Lacy}, {Kiman}, {Caselden},
  {Kirkpatrick}, {Meisner}, {Schneider}, {Kuchner}, {Bardalez Gagliuffi},
  {Beichman}, {Eisenhardt}, {Gelino}, {Gharib-Nezhad}, {Gonzales}, {Marocco},
  {Rothermich}, \& {Whiteford}}]{2024Natur.628..511F}
{Faherty}, J.~K., {Burningham}, B., {Gagne}, J., {et~al.} 2024, \nat, 628, 511,
  \dodoi{10.1038/s41586-024-07190-w}

\bibitem[{{Filippazzo}(2020)}]{2020ascl.soft11014F}
{Filippazzo}, J. 2020, {SEDkit: Spectral energy distribution construction and
  analysis tools}, Astrophysics Source Code Library, record ascl:2011.014

\bibitem[{{Filippazzo} {et~al.}(2015){Filippazzo}, {Rice}, {Faherty}, {Cruz},
  {Van Gordon}, \& {Looper}}]{2015ApJ...810..158F}
{Filippazzo}, J.~C., {Rice}, E.~L., {Faherty}, J., {et~al.} 2015, \apj, 810,
  158, \dodoi{10.1088/0004-637X/810/2/158}

\bibitem[{{Fontanive} {et~al.}(2018){Fontanive}, {Biller}, {Bonavita}, \&
  {Allers}}]{2018MNRAS.479.2702F}
{Fontanive}, C., {Biller}, B., {Bonavita}, M., \& {Allers}, K. 2018, \mnras,
  479, 2702, \dodoi{10.1093/mnras/sty1682}

\bibitem[{{Gardner} {et~al.}(2023){Gardner}, {Mather}, {Abbott}, {Abell},
  {Abernathy}, {Abney}, {Abraham}, {Abraham}, {Abul-Huda}, {Acton}, {Adams},
  {Adams}, {Adler}, {Adriaensen}, {Aguilar}, {Ahmed}, {Ahmed}, {Ahmed},
  {Albat}, {Albert}, {Alberts}, {Aldridge}, {Allen}, {Allen}, {Altenburg},
  {Altunc}, {Alvarez}, {{\'A}lvarez-M{\'a}rquez}, {Alves de Oliveira},
  {Ambrose}, {Anandakrishnan}, {Andersen}, {Anderson}, {Anderson}, {Anderson},
  {Anderson}, {Aprea}, {Archer}, {Arenberg}, {Argyriou}, {Arribas}, {Artigau},
  {Arvai}, {Atcheson}, {Atkinson}, {Averbukh}, {Aymergen}, {Bacinski},
  {Baggett}, {Bagnasco}, {Baker}, {Balzano}, {Banks}, {Baran}, {Barker},
  {Barrett}, {Barringer}, {Barto}, {Bast}, {Baudoz}, {Baum}, {Beatty},
  {Beaulieu}, {Bechtold}, {Beck}, {Beddard}, {Beichman}, {Bellagama}, {Bely},
  {Berger}, {Bergeron}, {Bernier}, {Bertch}, {Beskow}, {Betz}, {Biagetti},
  {Birkmann}, {Bjorklund}, {Blackwood}, {Blazek}, {Blossfeld}, {Bluth},
  {Boccaletti}, {Boegner}, {Bohlin}, {Boia}, {B{\"o}ker}, {Bonaventura},
  {Bond}, {Bosley}, {Boucarut}, {Bouchet}, {Bouwman}, {Bower}, {Bowers},
  {Bowers}, {Boyce}, {Boyer}, {Boyer}, {Boyer}, {Boyer}, {Bradley}, {Brady},
  {Brandl}, {Brannen}, {Breda}, {Bremmer}, {Brennan}, {Bresnahan}, {Bright},
  {Broiles}, {Bromenschenkel}, {Brooks}, {Brooks}, {Brown}, {Brown}, {Brown},
  {Bruce}, {Bryson}, {Bujanda}, {Bullock}, {Bunker}, {Bureo}, {Burt}, {Bush},
  {Bushouse}, {Bussman}, {Cabaud}, {Cale}, {Calhoon}, {Calvani}, {Canipe},
  {Caputo}, {Cara}, {Carey}, {Case}, {Cesari}, {Cetorelli}, {Chance},
  {Chandler}, {Chaney}, {Chapman}, {Charlot}, {Chayer}, {Cheezum}, {Chen},
  {Chen}, {Cherinka}, {Chichester}, {Chilton}, {Chittiraibalan}, {Clampin},
  {Clark}, {Clark}, {Clark}, {Claybrooks}, {Cleveland}, {Cohen}, {Cohen},
  {Col{\'o}n}, {Coleman}, {Colina}, {Comber}, {Comeau}, {Comer}, {Conde Reis},
  {Connolly}, {Conroy}, {Contos}, {Contreras}, {Cook}, {Cooper}, {Cooper},
  {Correia}, {Correnti}, {Cossou}, {Costanza}, {Coulais}, {Cox}, {Coyle},
  {Cracraft}, {Crew}, {Curtis}, {Cusveller}, {Da Costa Maciel}, {Dailey},
  {Daugeron}, {Davidson}, {Davies}, {Davis}, {Davis}, {Day}, {de Chambure}, {de
  Jong}, {De Marchi}, {Dean}, {Decker}, {Delisa}, {Dell}, \&
  {Dellagatta}}]{2023PASP..135f8001G}
{Gardner}, J.~P., {Mather}, J.~C., {Abbott}, R., {et~al.} 2023, \pasp, 135,
  068001, \dodoi{10.1088/1538-3873/acd1b5}

\bibitem[{{Jakobsen} {et~al.}(2022){Jakobsen}, {Ferruit}, {Alves de Oliveira},
  {Arribas}, {Bagnasco}, {Barho}, {Beck}, {Birkmann}, {B{\"o}ker}, {Bunker},
  {Charlot}, {de Jong}, {de Marchi}, {Ehrenwinkler}, {Falcolini}, {Fels},
  {Franx}, {Franz}, {Funke}, {Giardino}, {Gnata}, {Holota}, {Honnen}, {Jensen},
  {Jentsch}, {Johnson}, {Jollet}, {Karl}, {Kling}, {K{\"o}hler}, {Kolm},
  {Kumari}, {Lander}, {Lemke}, {L{\'o}pez-Caniego}, {L{\"u}tzgendorf},
  {Maiolino}, {Manjavacas}, {Marston}, {Maschmann}, {Maurer}, {Messerschmidt},
  {Moseley}, {Mosner}, {Mott}, {Muzerolle}, {Pirzkal}, {Pittet}, {Plitzke},
  {Posselt}, {Rapp}, {Rauscher}, {Rawle}, {Rix}, {R{\"o}del}, {Rumler},
  {Sabbi}, {Salvignol}, {Schmid}, {Sirianni}, {Smith}, {Strada}, {te Plate},
  {Valenti}, {Wettemann}, {Wiehe}, {Wiesmayer}, {Willott}, {Wright}, {Zeidler},
  \& {Zincke}}]{2022A&A...661A..80J}
{Jakobsen}, P., {Ferruit}, P., {Alves de Oliveira}, C., {et~al.} 2022, \aap,
  661, A80, \dodoi{10.1051/0004-6361/202142663}

\bibitem[{{Kirkpatrick} {et~al.}(2011){Kirkpatrick}, {Cushing}, {Gelino},
  {Griffith}, {Skrutskie}, {Marsh}, {Wright}, {Mainzer}, {Eisenhardt},
  {McLean}, {Thompson}, {Bauer}, {Benford}, {Bridge}, {Lake}, {Petty},
  {Stanford}, {Tsai}, {Bailey}, {Beichman}, {Bloom}, {Bochanski}, {Burgasser},
  {Capak}, {Cruz}, {Hinz}, {Kartaltepe}, {Knox}, {Manohar}, {Masters},
  {Morales-Calder{\'o}n}, {Prato}, {Rodigas}, {Salvato}, {Schurr}, {Scoville},
  {Simcoe}, {Stapelfeldt}, {Stern}, {Stock}, \& {Vacca}}]{2011ApJS..197...19K}
{Kirkpatrick}, J.~D., {Cushing}, M.~C., {Gelino}, C.~R., {et~al.} 2011, \apjs,
  197, 19, \dodoi{10.1088/0067-0049/197/2/19}

\bibitem[{{Kirkpatrick} {et~al.}(2012){Kirkpatrick}, {Gelino}, {Cushing},
  {Mace}, {Griffith}, {Skrutskie}, {Marsh}, {Wright}, {Eisenhardt}, {McLean},
  {Mainzer}, {Burgasser}, {Tinney}, {Parker}, \&
  {Salter}}]{2012ApJ...753..156K}
{Kirkpatrick}, J.~D., {Gelino}, C.~R., {Cushing}, M.~C., {et~al.} 2012, \apj,
  753, 156, \dodoi{10.1088/0004-637X/753/2/156}

\bibitem[{{Kirkpatrick} {et~al.}(2019){Kirkpatrick}, {Martin}, {Smart},
  {Cayago}, {Beichman}, {Marocco}, {Gelino}, {Faherty}, {Cushing}, {Schneider},
  {Mace}, {Tinney}, {Wright}, {Lowrance}, {Ingalls}, {Vrba}, {Munn}, {Dahm}, \&
  {McLean}}]{2019ApJS..240...19K}
{Kirkpatrick}, J.~D., {Martin}, E.~C., {Smart}, R.~L., {et~al.} 2019, \apjs,
  240, 19, \dodoi{10.3847/1538-4365/aaf6af}

\bibitem[{{Kirkpatrick} {et~al.}(2021){Kirkpatrick}, {Gelino}, {Faherty},
  {Meisner}, {Caselden}, {Schneider}, {Marocco}, {Cayago}, {Smart},
  {Eisenhardt}, {Kuchner}, {Wright}, {Cushing}, {Allers}, {Bardalez Gagliuffi},
  {Burgasser}, {Gagn{\'e}}, {Logsdon}, {Martin}, {Ingalls}, {Lowrance},
  {Abrahams}, {Aganze}, {Gerasimov}, {Gonzales}, {Hsu}, {Kamraj}, {Kiman},
  {Rees}, {Theissen}, {Ammar}, {Andersen}, {Beaulieu}, {Colin}, {Elachi},
  {Goodman}, {Gramaize}, {Hamlet}, {Hong}, {Jonkeren}, {Khalil}, {Martin},
  {Pendrill}, {Pumphrey}, {Rothermich}, {Sainio}, {Stenner}, {Tanner},
  {Th{\'e}venot}, {Voloshin}, {Walla}, {W{\k{e}}dracki}, \& {Backyard Worlds:
  Planet 9 Collaboration}}]{2021ApJS..253....7K}
{Kirkpatrick}, J.~D., {Gelino}, C.~R., {Faherty}, J.~K., {et~al.} 2021, \apjs,
  253, 7, \dodoi{10.3847/1538-4365/abd107}

\bibitem[{{Kirkpatrick} {et~al.}(2024){Kirkpatrick}, {Marocco}, {Gelino},
  {Raghu}, {Faherty}, {Bardalez Gagliuffi}, {Schurr}, {Apps}, {Schneider},
  {Meisner}, {Kuchner}, {Caselden}, {Smart}, {Casewell}, {Raddi}, {Kesseli},
  {Stevnbak Andersen}, {Antonini}, {Beaulieu}, {Bickle}, {Bilsing}, {Chieng},
  {Colin}, {Deen}, {Dereveanco}, {Doll}, {Durantini Luca}, {Frazer}, {Gantier},
  {Gramaize}, {Grant}, {Hamlet}, {Higashimura}, {Hyogo}, {Ja{\l}owiczor},
  {Jonkeren}, {Kabatnik}, {Kiwy}, {Martin}, {Michaels}, {Pendrill}, {Pessanha
  Machado}, {Pumphrey}, {Rothermich}, {Russwurm}, {Sainio}, {Sanchez},
  {Sapelkin-Tambling}, {Sch{\"u}mann}, {Selg-Mann}, {Singh}, {Stenner}, {Sun},
  {Tanner}, {Th{\'e}venot}, {Ventura}, {Voloshin}, {Walla}, {W{\k{e}}dracki},
  {Adorno}, {Aganze}, {Allers}, {Brooks}, {Burgasser}, {Calamari}, {Connor},
  {Costa}, {Eisenhardt}, {Gagn{\'e}}, {Gerasimov}, {Gonzales}, {Hsu}, {Kiman},
  {Li}, {Low}, {Mamajek}, {Pantoja}, {Popinchalk}, {Rees}, {Stern},
  {Su{\'a}rez}, {Theissen}, {Tsai}, {Vos}, {Zurek}, \& {The Backyard Worlds:
  Planet 9 Collaboration}}]{2024ApJS..271...55K}
{Kirkpatrick}, J.~D., {Marocco}, F., {Gelino}, C.~R., {et~al.} 2024, \apjs,
  271, 55, \dodoi{10.3847/1538-4365/ad24e2}

\bibitem[{{Knapp} {et~al.}(2004){Knapp}, {Leggett}, {Fan}, {Marley}, {Geballe},
  {Golimowski}, {Finkbeiner}, {Gunn}, {Hennawi}, {Ivezi{\'c}}, {Lupton},
  {Schlegel}, {Strauss}, {Tsvetanov}, {Chiu}, {Hoversten}, {Glazebrook},
  {Zheng}, {Hendrickson}, {Williams}, {Uomoto}, {Vrba}, {Henden}, {Luginbuhl},
  {Guetter}, {Munn}, {Canzian}, {Schneider}, \&
  {Brinkmann}}]{2004AJ....127.3553K}
{Knapp}, G.~R., {Leggett}, S.~K., {Fan}, X., {et~al.} 2004, \aj, 127, 3553,
  \dodoi{10.1086/420707}

\bibitem[{{Lacy} \& {Burrows}(2023)}]{2023ApJ...950....8L}
{Lacy}, B., \& {Burrows}, A. 2023, \apj, 950, 8,
  \dodoi{10.3847/1538-4357/acc8cb}

\bibitem[{{Leggett} {et~al.}(2015){Leggett}, {Morley}, {Marley}, \&
  {Saumon}}]{2015ApJ...799...37L}
{Leggett}, S.~K., {Morley}, C.~V., {Marley}, M.~S., \& {Saumon}, D. 2015, \apj,
  799, 37, \dodoi{10.1088/0004-637X/799/1/37}

\bibitem[{{Leggett} {et~al.}(2013){Leggett}, {Morley}, {Marley}, {Saumon},
  {Fortney}, \& {Visscher}}]{2013ApJ...763..130L}
{Leggett}, S.~K., {Morley}, C.~V., {Marley}, M.~S., {et~al.} 2013, \apj, 763,
  130, \dodoi{10.1088/0004-637X/763/2/130}

\bibitem[{{Leggett} {et~al.}(2012){Leggett}, {Saumon}, {Marley}, {Lodders},
  {Canty}, {Lucas}, {Smart}, {Tinney}, {Homeier}, {Allard}, {Burningham},
  {Day-Jones}, {Fegley}, {Ishii}, {Jones}, {Marocco}, {Pinfield}, \&
  {Tamura}}]{2012ApJ...748...74L}
{Leggett}, S.~K., {Saumon}, D., {Marley}, M.~S., {et~al.} 2012, \apj, 748, 74,
  \dodoi{10.1088/0004-637X/748/2/74}

\bibitem[{{Leggett} {et~al.}(2021){Leggett}, {Tremblin}, {Phillips}, {Dupuy},
  {Marley}, {Morley}, {Schneider}, {Caselden}, {Guillaume}, \&
  {Logsdon}}]{2021ApJ...918...11L}
{Leggett}, S.~K., {Tremblin}, P., {Phillips}, M.~W., {et~al.} 2021, \apj, 918,
  11, \dodoi{10.3847/1538-4357/ac0cfe}

\bibitem[{{Liu} {et~al.}(2012){Liu}, {Dupuy}, {Bowler}, {Leggett}, \&
  {Best}}]{2012ApJ...758...57L}
{Liu}, M.~C., {Dupuy}, T.~J., {Bowler}, B.~P., {Leggett}, S.~K., \& {Best},
  W.~M.~J. 2012, \apj, 758, 57, \dodoi{10.1088/0004-637X/758/1/57}

\bibitem[{{Lodders} \& {Fegley}(2002)}]{2002Icar..155..393L}
{Lodders}, K., \& {Fegley}, B. 2002, \icarus, 155, 393,
  \dodoi{10.1006/icar.2001.6740}

\bibitem[{{Lucas} {et~al.}(2010){Lucas}, {Tinney}, {Burningham}, {Leggett},
  {Pinfield}, {Smart}, {Jones}, {Marocco}, {Barber}, {Yurchenko}, {Tennyson},
  {Ishii}, {Tamura}, {Day-Jones}, {Adamson}, {Allard}, \&
  {Homeier}}]{2010MNRAS.408L..56L}
{Lucas}, P.~W., {Tinney}, C.~G., {Burningham}, B., {et~al.} 2010, \mnras, 408,
  L56, \dodoi{10.1111/j.1745-3933.2010.00927.x}

\bibitem[{{Luhman} {et~al.}(2024){Luhman}, {Tremblin}, {Alves de Oliveira},
  {Birkmann}, {Baraffe}, {Chabrier}, {Manjavacas}, {Parker}, \&
  {Valenti}}]{2024AJ....167....5L}
{Luhman}, K.~L., {Tremblin}, P., {Alves de Oliveira}, C., {et~al.} 2024, \aj,
  167, 5, \dodoi{10.3847/1538-3881/ad0b72}

\bibitem[{{Mainzer} {et~al.}(2023){Mainzer}, {Masiero}, {Abell}, {Bauer},
  {Bottke}, {Buratti}, {Carey}, {Cotto-Figueroa}, {Cutri}, {Dahlen},
  {Eisenhardt}, {Fernandez}, {Furfaro}, {Grav}, {Hoffman}, {Kelley}, {Kim},
  {Kirkpatrick}, {Lawler}, {Lilly}, {Liu}, {Marocco}, {Marsh}, {Masci},
  {McMurtry}, {Pourrahmani}, {Reinhart}, {Ressler}, {Satpathy}, {Schambeau},
  {Sonnett}, {Spahr}, {Surace}, {Vaquero}, {Wright}, {Zengilowski}, \& {NEO
  Surveyor Mission Team}}]{2023PSJ.....4..224M}
{Mainzer}, A.~K., {Masiero}, J.~R., {Abell}, P.~A., {et~al.} 2023, \psj, 4,
  224, \dodoi{10.3847/PSJ/ad0468}

\bibitem[{{Marley} {et~al.}(2021){Marley}, {Saumon}, {Visscher}, {Lupu},
  {Freedman}, {Morley}, {Fortney}, {Seay}, {Smith}, {Teal}, \&
  {Wang}}]{2021ApJ...920...85M}
{Marley}, M.~S., {Saumon}, D., {Visscher}, C., {et~al.} 2021, \apj, 920, 85,
  \dodoi{10.3847/1538-4357/ac141d}

\bibitem[{{Martin} {et~al.}(2018){Martin}, {Kirkpatrick}, {Beichman}, {Smart},
  {Faherty}, {Gelino}, {Cushing}, {Schneider}, {Wright}, {Lowrance}, {Ingalls},
  {Tinney}, {McLean}, {Logsdon}, \& {Lebreton}}]{2018ApJ...867..109M}
{Martin}, E.~C., {Kirkpatrick}, J.~D., {Beichman}, C.~A., {et~al.} 2018, \apj,
  867, 109, \dodoi{10.3847/1538-4357/aae1af}

\bibitem[{{Matthews} {et~al.}(2024){Matthews}, {Carter}, {Pathak}, {Morley},
  {Phillips}, {P.~M.}, {Feng}, {Bonse}, {Boogaard}, {Burt}, {Crossfield},
  {Douglas}, {Henning}, {Hom}, {Ko}, {Kasper}, {Lagrange}, {Petit dit de la
  Roche}, \& {Philipot}}]{2024Natur.633..789M}
{Matthews}, E.~C., {Carter}, A.~L., {Pathak}, P., {et~al.} 2024, \nat, 633,
  789, \dodoi{10.1038/s41586-024-07837-8}

\bibitem[{{Miles} {et~al.}(2020){Miles}, {Skemer}, {Morley}, {Marley},
  {Fortney}, {Allers}, {Faherty}, {Geballe}, {Visscher}, {Schneider}, {Lupu},
  {Freedman}, \& {Bjoraker}}]{2020AJ....160...63M}
{Miles}, B.~E., {Skemer}, A. J.~I., {Morley}, C.~V., {et~al.} 2020, \aj, 160,
  63, \dodoi{10.3847/1538-3881/ab9114}

\bibitem[{{Morley} {et~al.}(2019){Morley}, {Skemer}, {Miles}, {Line}, {Lopez},
  {Brogi}, {Freedman}, \& {Marley}}]{2019ApJ...882L..29M}
{Morley}, C.~V., {Skemer}, A.~J., {Miles}, B.~E., {et~al.} 2019, \apjl, 882,
  L29, \dodoi{10.3847/2041-8213/ab3c65}

\bibitem[{{Mukherjee} {et~al.}(2022){Mukherjee}, {Fortney}, {Batalha},
  {Karalidi}, {Marley}, {Visscher}, {Miles}, \& {Skemer}}]{2022ApJ...938..107M}
{Mukherjee}, S., {Fortney}, J.~J., {Batalha}, N.~E., {et~al.} 2022, \apj, 938,
  107, \dodoi{10.3847/1538-4357/ac8dfb}

\bibitem[{{Offner} {et~al.}(2022){Offner}, {Moe}, {Kratter}, {Sadavoy},
  {Jensen}, \& {Tobin}}]{2022arXiv220310066O}
{Offner}, S. S.~R., {Moe}, M., {Kratter}, K.~M., {et~al.} 2022, arXiv e-prints,
  arXiv:2203.10066, \dodoi{10.48550/arXiv.2203.10066}

\bibitem[{{Opitz} {et~al.}(2016){Opitz}, {Tinney}, {Faherty}, {Sweet},
  {Gelino}, \& {Kirkpatrick}}]{2016ApJ...819...17O}
{Opitz}, D., {Tinney}, C.~G., {Faherty}, J.~K., {et~al.} 2016, \apj, 819, 17,
  \dodoi{10.3847/0004-637X/819/1/17}

\bibitem[{{Raghavan} {et~al.}(2010){Raghavan}, {McAlister}, {Henry}, {Latham},
  {Marcy}, {Mason}, {Gies}, {White}, \& {ten Brummelaar}}]{2010ApJS..190....1R}
{Raghavan}, D., {McAlister}, H.~A., {Henry}, T.~J., {et~al.} 2010, \apjs, 190,
  1, \dodoi{10.1088/0067-0049/190/1/1}

\bibitem[{{Rieke} {et~al.}(2023){Rieke}, {Kelly}, {Misselt}, {Stansberry},
  {Boyer}, {Beatty}, {Egami}, {Florian}, {Greene}, {Hainline}, {Leisenring},
  {Roellig}, {Schlawin}, {Sun}, {Tinnin}, {Williams}, {Willmer}, {Wilson},
  {Clark}, {Rohrbach}, {Brooks}, {Canipe}, {Correnti}, {DiFelice}, {Gennaro},
  {Girard}, {Hartig}, {Hilbert}, {Koekemoer}, {Nikolov}, {Pirzkal}, {Rest},
  {Robberto}, {Sunnquist}, {Telfer}, {Wu}, {Ferry}, {Lewis}, {Baum},
  {Beichman}, {Doyon}, {Dressler}, {Eisenstein}, {Ferrarese}, {Hodapp},
  {Horner}, {Jaffe}, {Johnstone}, {Krist}, {Martin}, {McCarthy}, {Meyer},
  {Rieke}, {Trauger}, \& {Young}}]{2023PASP..135b8001R}
{Rieke}, M.~J., {Kelly}, D.~M., {Misselt}, K., {et~al.} 2023, \pasp, 135,
  028001, \dodoi{10.1088/1538-3873/acac53}

\bibitem[{{Schneider} {et~al.}(2015){Schneider}, {Cushing}, {Kirkpatrick},
  {Gelino}, {Mace}, {Wright}, {Eisenhardt}, {Skrutskie}, {Griffith}, \&
  {Marsh}}]{2015ApJ...804...92S}
{Schneider}, A.~C., {Cushing}, M.~C., {Kirkpatrick}, J.~D., {et~al.} 2015,
  \apj, 804, 92, \dodoi{10.1088/0004-637X/804/2/92}

\bibitem[{{Skemer} {et~al.}(2016){Skemer}, {Morley}, {Allers}, {Geballe},
  {Marley}, {Fortney}, {Faherty}, {Bjoraker}, \& {Lupu}}]{2016ApJ...826L..17S}
{Skemer}, A.~J., {Morley}, C.~V., {Allers}, K.~N., {et~al.} 2016, \apjl, 826,
  L17, \dodoi{10.3847/2041-8205/826/2/L17}

\bibitem[{{Su{\'a}rez} \& {Metchev}(2022)}]{2022MNRAS.513.5701S}
{Su{\'a}rez}, G., \& {Metchev}, S. 2022, \mnras, 513, 5701,
  \dodoi{10.1093/mnras/stac1205}

\bibitem[{{Su{\'a}rez} {et~al.}(2021){Su{\'a}rez}, {Metchev}, {Leggett},
  {Saumon}, \& {Marley}}]{2021ApJ...920...99S}
{Su{\'a}rez}, G., {Metchev}, S., {Leggett}, S.~K., {Saumon}, D., \& {Marley},
  M.~S. 2021, \apj, 920, 99, \dodoi{10.3847/1538-4357/ac1418}

\bibitem[{{Tsuji} {et~al.}(1996){Tsuji}, {Ohnaka}, \&
  {Aoki}}]{1996AandA...305L...1T}
{Tsuji}, T., {Ohnaka}, K., \& {Aoki}, W. 1996, \aap, 305, L1

\bibitem[{{Winters} {et~al.}(2019){Winters}, {Henry}, {Jao}, {Subasavage},
  {Chatelain}, {Slatten}, {Riedel}, {Silverstein}, \&
  {Payne}}]{2019AJ....157..216W}
{Winters}, J.~G., {Henry}, T.~J., {Jao}, W.-C., {et~al.} 2019, \aj, 157, 216,
  \dodoi{10.3847/1538-3881/ab05dc}

\bibitem[{{Zahnle} \& {Marley}(2014)}]{2014ApJ...797...41Z}
{Zahnle}, K.~J., \& {Marley}, M.~S. 2014, \apj, 797, 41,
  \dodoi{10.1088/0004-637X/797/1/41}

\bibitem[{{Zhang} {et~al.}(2021){Zhang}, {Liu}, {Claytor}, {Best}, {Dupuy}, \&
  {Siverd}}]{2021ApJ...916L..11Z}
{Zhang}, Z., {Liu}, M.~C., {Claytor}, Z.~R., {et~al.} 2021, \apjl, 916, L11,
  \dodoi{10.3847/2041-8213/ac1123}

\end{thebibliography}
\bibliographystyle{aasjournal}

\end{document}